\documentclass[preprint,12pt]{elsarticle}

\usepackage{amssymb}
\usepackage{soul,color}
\usepackage{amsmath}
\usepackage{array}
\usepackage{booktabs}   % For formal tables
\usepackage{tablefootnote}
\usepackage[table]{xcolor}%
\usepackage{graphicx}
\usepackage{subcaption}
\usepackage{float} 
\usepackage{placeins} % for [H]
\usepackage{comment}
\graphicspath{{figure/}} 

\usepackage[margin=1.25in]{geometry}

\DeclareMathOperator*{\argmin}{arg\,min}

\journal{Journal of Energy Storage}

\begin{document}

\begin{frontmatter}

%% Title, authors and addresses

%% use the tnoteref command within \title for footnotes;
%% use the tnotetext command for theassociated footnote;
%% use the fnref command within \author or \address for footnotes;
%% use the fntext command for theassociated footnote;
%% use the corref command within \author for corresponding author footnotes;
%% use the cortext command for theassociated footnote;
%% use the ead command for the email address,
%% and the form \ead[url] for the home page:
%% \title{Title\tnoteref{label1}}
%% \tnotetext[label1]{}
%% \author{Name\corref{cor1}\fnref{label2}}
%% \ead{email address}
%% \ead[url]{home page}
%% \fntext[label2]{}
%% \cortext[cor1]{}
%% \affiliation{organization={},
%%             addressline={},
%%             city={},
%%             postcode={},
%%             state={},
%%             country={}}
%% \fntext[label3]{}

\title{PINEAPPLE: Physics-Informed Neuro-Evolution Algorithm for Prognostic Parameter Inference in Lithium-Ion Battery Electrodes}

%% use optional labels to link authors explicitly to addresses:
%% \author[label1,label2]{}
%% \affiliation[label1]{organization={},
%%             addressline={},
%%             city={},
%%             postcode={},
%%             state={},
%%             country={}}
%%
%% \affiliation[label2]{organization={},
%%             addressline={},
%%             city={},
%%             postcode={},
%%             state={},
%%             country={}}

%\begin{comment}

\author[inst1]{Karkulali Pugalenthi\fnref{label1}}
\author[inst2]{Jian Cheng Wong\fnref{label1}}
\author[inst3]{Qizheng Yang}
\author[inst2]{Pao-Hsiung Chiu}
\author[inst4]{My Ha Dao}
\author[inst5]{Nagarajan Raghavan}
\author[inst2,inst6]{Chinchun Ooi\fnref{label2}}

\fntext[label1]{These authors contributed equally to this work.}
\fntext[label2]{Corresponding author: ooicc@a-star.edu.sg}

\affiliation[inst1]{organization={Singapore Institute of Manufacturing Technology, Agency for Science, Technology and Research},%Department and Organization
            addressline={5 Cleantech Loop}, 
            city={Singapore},
            postcode={636732},
           % state={State One},
            country={Singapore}}
            
\affiliation[inst2]{organization={Institute of High Performance Computing, Agency for Science, Technology and Research},%Department and Organization
            addressline={1 Fusionopolis Way}, 
            city={Singapore},
            postcode={138632}, 
            %state={Singapore},
            country={Singapore}}

\affiliation[inst3]{organization={Department of Mathematics, The Chinese University of Hong Kong},
            addressline={Shatin, N.T.}, 
            city={Hong Kong SAR},
            %postcode={138632}, 
            %state={Singapore},
            country={China}}

\affiliation[inst4]{organization={Technology Centre for Offshore and Marine, Singapore},%Department and Organization
            addressline={12 Prince George’s Park}, 
            city={Singapore},
            postcode={118411}, 
            %state={Singapore},
            country={Singapore}}

\affiliation[inst5]{organization={Engineering Product Design Pillar, Singapore University of Technology and Design},%Department and Organization
            addressline={8 Somapah Road}, 
            city={Singapore},
            postcode={487372}, 
            %state={Singapore},
            country={Singapore}}
            
\affiliation[inst6]{organization={Centre for Frontier AI Research, Agency for Science, Technology and Research},%Department and Organization
            addressline={1 Fusionopolis Way}, 
            city={Singapore},
            postcode={138632}, 
            %state={Singapore},
            country={Singapore}}

%\end{comment}

\begin{abstract}
%% Text of abstract
Accurate, real-time, yet non-destructive estimation of internal states in lithium-ion batteries is critical for predicting degradation, optimizing usage strategies, and extending operational lifespan. While physics-based models, such as the single-particle model, utilize fundamental mechanistic insights to link macro-scale measurements (e.g., voltage response) to batteries' internal electrochemical dynamics, their relatively high computational cost limits real-time diagnostics and continuous inference during battery pack operation. Here, we introduce PINEAPPLE (Physics-Informed Neuro-Evolution Algorithm for Prognostic Parameter inference in Lithium-ion battery Electrodes), a novel framework that integrates physics-informed neural networks (PINNs) with an evolutionary search algorithm to enable rapid, scalable, and interpretable parameter inference with potential for application to next-generation batteries. % computationally efficient parameter inference from voltage-time discharge curves.
The meta-learned PINN utilizes fundamental physics principles to achieve accurate zero-shot prediction of electrode behavior with test errors below 0.1$\%$ while maintaining an order-of-magnitude speed-up over conventional solvers. PINEAPPLE demonstrates robust parameter inference solely from voltage-time discharge curves across multiple batteries from the open-source CALCE repository, recovering the evolution of key internal state parameters such as Li-ion diffusion coefficients across usage cycles. Notably, the inferred cycle-dependent evolution of these parameters exhibit consistent trends across different batteries without any customized degradation physics-embedded heuristic, highlighting the effective regularizing effect and robustness that can be conferred through incorporation of fundamental physics in PINEAPPLE. By enabling computationally efficient, real-time parameter estimation, PINEAPPLE offers a promising route towards the non-destructive, physics-based characterization of inter-cell and intra-cell variability of battery modules and battery packs, thereby unlocking new opportunities for downstream on-the-fly needs in next-generation battery management systems such as individual cell-scale state-of-health diagnostics.
%Battery informatics is an emerging field that combines physics-based models with AI-driven analytics to accelerate the design and discovery of novel battery materials. Machine learning (ML)-based surrogate models have emerged as fast and computationally efficient alternatives to traditional first-principles calculations, enabling the prediction of critical properties such as mechanical characteristics and transport phenomena like diffusion coefficients in electrode materials. While conventional methods are resource-intensive and require significant domain expertise, we propose a physics-informed neural network (PINN) surrogate model to replace the physics-based single particle model for deducing internal battery parameters from voltage discharge curves. Our developed PINN surrogate model is pretrained using the Baldwinian evolution approach, leveraging voltage-time (V-t) curves generated with PyBAMM. This pretrained PINN model is further utilized for parameter inference using the Covariance Matrix Adaptation Evolution Strategy (CMA-ES) on open-source datasets from the CALCE repository. The model efficiently deduces key parameters such as the Li-ion diffusion coefficient and molar flux, providing critical insights into battery behaviour. The proposed PINN model offers a computationally efficient and accurate alternative for accelerating the development of next-generation battery chemistries and improving state-of-health (SoH) diagnostics, particularly in scenarios with limited information about electrode properties.
\end{abstract}

%%Graphical abstract
%\begin{graphicalabstract}
%\includegraphics{figure/grabs.pdf}
%\end{graphicalabstract}

%%Research highlights
%\begin{highlights}
%\item Research highlight 1
%\item Research highlight 2
%\end{highlights}

\begin{keyword}
%% keywords here, in the form: keyword \sep keyword
lithium-ion battery \sep physics-informed neural networks \sep evolutionary algorithm \sep meta-learning \sep inverse inference
%% PACS codes here, in the form: \PACS code \sep code
%\PACS 0000 \sep 1111
%% MSC codes here, in the form: \MSC code \sep code
%% or \MSC[2008] code \sep code (2000 is the default)
%\MSC 0000 \sep 1111
\end{keyword}

\end{frontmatter}

%% \linenumbers

%% main text
\section{Introduction} \label{sec:introduction}

Innovations in energy storage technology are fundamental to a more sustainable future, powering trends from electrification of transportation to the seamless integration of intermittent renewable energy sources such as solar and wind into the modern power grid. Lithium-ion batteries (Li-ion) are the current most important candidate technology at the heart of this transition. For example, the Hornsdale Power Reserve in South Australia is a key demonstration of how large-scale storage can stabilize grid frequency and reduce reliance on fossil fuels~\cite{reserve2018hornsdale}.

Given recent improvements in battery capacity and production cost, effective battery management systems (BMS) are increasingly seen as the next critical frontier for safer and more sustainable operational implementation. For example, innovations in BMS enable optimized fast-charging which extend the cycle life of the batteries by 40\% ~\cite{attia2020closed}, while simultaneously being imperative for mitigating thermal runaway, a potential cause of significant hazard as exemplified by electric vehicle fires and similar incidents in aviation ~\cite{feng2018thermal, al2020thermal}. Hence, the development and incorporation of methodologies for accurately predicting the state of health (SoH) and remaining useful life (RUL) of batteries in BMS have drawn significant attention for their anticipated benefits to battery efficiency, longevity and safety. However, current systems still face challenges in non-destructive, real-time battery internal state inference.

Existing approaches to battery health estimation utilize either physics-based or data-driven methods, each with their own pros and cons~\cite{downey2019physics, cadini2019state}. Purely data-driven methods utilize methods such as deep learning to predict SoH or RUL from macro-scale measurements such as voltage discharge curves and are prized for their ability to directly extract valuable insights from real-world data. Recent advances include gated recurrent and state-space inspired sequence models for long-horizon time-series forecasting ~\cite{li2025channel}, adaptive gated RNNs with differential information storage mechanisms for RUL predictions ~\cite{xiang2024single}, the use of principal component-based feature engineering in state-of-charge estimation ~\cite{mehta2024optimized}, and lightweight deformable neural networks designed for efficient lithium-ion battery prognostics ~\cite{wu2025ldnet}. While their predictive performance can be good, their robustness varies with quality and comprehensiveness of the data, which is problematic given the complexity of these electrochemical systems, wide range of operational possibilities, and real-world noise and uncertainty. Critically, internal physical parameters such as electrode material properties and key kinetic and transport characteristics which are known to be mechanistically important are frequently excluded from these models due to challenges in direct measurement or inference during operation~\cite{kouhestani2023data}.

Physics-based methods provide a complementary approach to diagnosing battery health through a mechanistic model of the underlying physical and electrochemical processes driving degradation, especially with advances in understanding and modelling the relation between battery failure and underlying mechanisms. Physics-based electrochemical models such as the Single Particle Model (SPM) ~\cite{gopalakrishnan2021composite, han2015simplification, li2018single} and the Doyle-Fuller-Newman (DFN) ~\cite{uddin2016characterising, xu2022enabling, bizeray2016state} model are widely used to characterize the chemical and mechanical processes that occur during charge and discharge cycles. These models can capture important phenomena like lithium-ion diffusion, interfacial reactions, and structural changes in the electrodes, making them suited for understanding battery degradation at a fundamental level. For example, monotonic trends observed in capacity fade and voltage drop have been shown to be driven by several underlying failure mechanisms, including electrode degradation ~\cite{pender2020electrode}, lithium plating ~\cite{lin2021lithium, von2019modeling}, electrode cracking ~\cite{boyce2022cracking}, electrolyte decomposition ~\cite{rinkel2022two}, and structural changes ~\cite{hu2018probing}. Despite their accuracy and interpretability, electrochemical models require extensive enumeration of key parameters, such as diffusion coefficient, internal resistance, and reaction kinetics, when applied in the real-world. This can be challenging as these key characteristics vary from battery to battery and cannot be easily inferred from measurable outputs like voltage or capacity, necessitating destructive tear-downs of the battery. In addition, these models can be computationally intensive, limiting their applicability to real-time onboard implementation. Simpler equivalent circuit models (ECMs) have been proposed as a more physically interpretable and computationally efficient framework, but remain primarily phenomenological, with a similar need for proper parameter calibration, and limited generalizability under novel operating conditions ~\cite{xu2022co, wang2022electrochemical,hosseininasab2022state, sepasiahooyi2024fault}.

To address these limitations, Physics-Informed Neural Networks (PINNs) offer a promising hybrid framework that integrates the strengths of physics-based modeling with the flexibility of data-driven methods~\cite{nascimento2021hybrid}. By embedding underlying physical laws or physics-informed constraints such as monotonicity and degradation dynamics into the model, PINNs enable more physical and accurate battery diagnosis and prognosis. Multiple authors have previously developed physics-informed frameworks with empirical degradation models for predicting state of health and battery management, including across different battery chemistries~\cite{wen2023physics, sun2025method, wang2024physics, wang2023inherently, wang2024physical, lin2025lightweight}. However, these models focus primarily on SoH estimation, and do not tackle the more challenging inverse problem of inferring unobservable internal state parameters within the battery which can be beneficial for mechanistic interpretability. For example, monitoring of (inferred) lithium-ion diffusion rates through electrodes can be a more mechanistically robust and interpretable indicator of SoH than typical phenomenological or empirical SoH models as reduction in lithium-ion diffusion is known to be related to key failure mechanisms such as lithium plating and electrode degradation. Electrode-level diagnostics which track key parameters such as total capacity, individual electrode capacities, and lithium inventory capacity may be able to provide deeper insight into internal degradation~\cite{xiong2025enhanced, singh2023hybrid}.

Separately, researchers have demonstrated the ability for PINNs to solve the electrochemical models, e.g., single particle models (SPM) and pseudo-two-dimensional (P2D), with significant model acceleration~\cite{mendez2024physics, hassanaly2024pinn, hassanaly2024pinn1}. This acceleration was shown to be advantageous for Bayesian calibration of the battery parameters~\cite{hassanaly2024pinn1}. However, the primary focus of prior work was on demonstrating acceleration (on simulated model data) rather than real-world battery parameter inference. Moreover, none of the aforementioned work demonstrated the application of an end-to-end PINN-enabled framework to the extraction of \textit{evolutionary trends in key electrochemical parameters} such as the diffusion coefficients of the electrodes, molar flux, and maximum lithium concentration in real-world battery cycling. Critically, this inverse problem is potentially ill-posed, where numerous combinations of internal parameters can produce deceptively similar voltage-time measurements (especially in the presence of noise), demanding a highly robust and efficient inference methodology.

We address this critical gap through the PINEAPPLE framework (Physics-Informed Neuro-Evolution Algorithm for Prognostic Parameter Inference in Lithium-ion battery Electrodes) that is designed to enable tracking of the \textit{cycle-dependent evolution} of fundamental battery internal state parameters such as diffusion coefficient, thereby offering deeper interpretability and enhanced predictive capabilities for Li-ion batteries. We leverage a Baldwinian neuro-evolution strategy to meta-learn a highly generalizable PINN~\cite{wong2023generalizable} that can achieve zero-shot prediction of electrode behavior based on the SPM across a wide range of physical parameters while being an order-of-magnitude faster than a conventional numerical solver. The PINN model is further integrated with an evolutionary search algorithm to enable rapid and robust inference of key physical parameters within a mechanistic model solely from easily accessible voltage-time (V-t) measurements. In particular, evolutionary algorithms are gradient-free, and can provide a population-based estimation of the parameters, mitigating the ill-posed nature of the problem. This shift from model acceleration to real-time parameter inference positions PINEAPPLE as a practical and easily deployable framework for next generation BMS applications. %PINEAPPLE is a novel framework that integrates a meta-learned PINN obtained through neuro-evolution with an evolutionary search algorithm to enable rapid and robust inference of key phsyical parameters solely from readily-available voltage-time (V-t) curves. 

The open-source CALCE repository provides limited information on battery electrode materials and other critical electrochemical parameters. Nonetheless, we demonstrate that PINEAPPLE can track the evolution of parameters, such as the Li-ion diffusion coefficients in both the anode and cathode, across experimental data obtained from CALCE. Crucially, this hybrid approach utilizes a fundamental physical law (Fick's Law of Diffusion) while obviating the need for detailed material-specific information or explicit determination of every model parameter or real-world inter-battery variability by shifting the focus to the relative evolution of key physical quantities such as diffusion coefficients. Our results reveal that these inferred parameters exhibit consistent, physically plausible trends, strongly correlating with capacity fade across cycles, and offering a window into the dominant degradation mechanisms at different stages. While the current results are still at the proof-of-concept stage, and require future demonstration across more varied battery chemistries, operating profiles, and aging conditions, PINEAPPLE is a promising and scalable framework that offers a new path towards more interpretable and physics-aware diagnostics for next-generation BMS through a computationally efficient, real-time window into the battery's internal physical state.

The main contributions of the present work are as follows:
\begin{enumerate}
    \item We introduce PINEAPPLE, a physics-informed neuro-evolution framework that combines meta-learned PINNs with evolutionary search for rapid, non-destructive inference of internal electrochemical state parameters in lithium-ion batteries from voltage–time measurements alone.
    \item We employ a Baldwinian meta-learning strategy to train a highly generalizable PINN model for the Single Particle Model (SPM) that can achieve zero-shot, physics-consistent prediction of electrode lithium-ion concentration dynamics with 10 $\times$ speed-up over PyBAMM at similar accuracies.
    \item We formulate the non-destructive inverse inference of internal state parameters from V-t data across cycles as an evolutionary search over physically meaningful scaling factors, enabling fast and robust parameter inference without requiring battery-specific calibration.
    \item We demonstrate the ability of PINEAPPLE to recover consistent and physically-interpretable degradation trends in key internal parameters such as effective anode and cathode diffusion coefficients using real-world data from the open-source CALCE dataset.
\end{enumerate}

The remainder of this manuscript is organized as follows. Section~2 introduces the real-world CALCE battery dataset used in this study, while Section~3 presents the underlying first-principles model used. Section~4 details the proposed PINEAPPLE framework, comprising the offline meta-learning of a generalizable PINN model for lithium-ion batteries (Section~4.2) and the online inference of cycle-dependent internal state parameters based on voltage–time discharge curves (Section~4.3). Section~5 presents comprehensive experimental results on both synthetic and real-world battery data, including quantitative evaluation of inference speed and accuracy and mechanistic analysis of inferred degradation trends. Finally, Section~6 concludes with a brief summary of key findings and limitations of the present study, and outlines promising directions for future extensions. \\

\section{Real-world dataset} \label{sec:dataset}

Several lithium-ion battery research teams have provided open access to their experimental test data on different types of Li-ion batteries operating under diverse operating conditions~\cite{dos2021lithium}. These datasets are widely utilized for degradation modeling, including battery state estimation, remaining useful life prediction, and reliability analysis, and generally comprise current or voltage measurements obtained across repeated charge and discharge cycles in the laboratory. 

%%%%%%%%%%%%%%%%%%%%%%%%%%%%%%%%%%%%%%%%%%%%%%%%%%%% FIG %%%%%%%%%%%%%%%%%%%%%%%%%%%%%%%%%%%%%%%%%%%%%%%%%%%%
\begin{figure}[!hbt]
\centering
\includegraphics[width=1\linewidth]{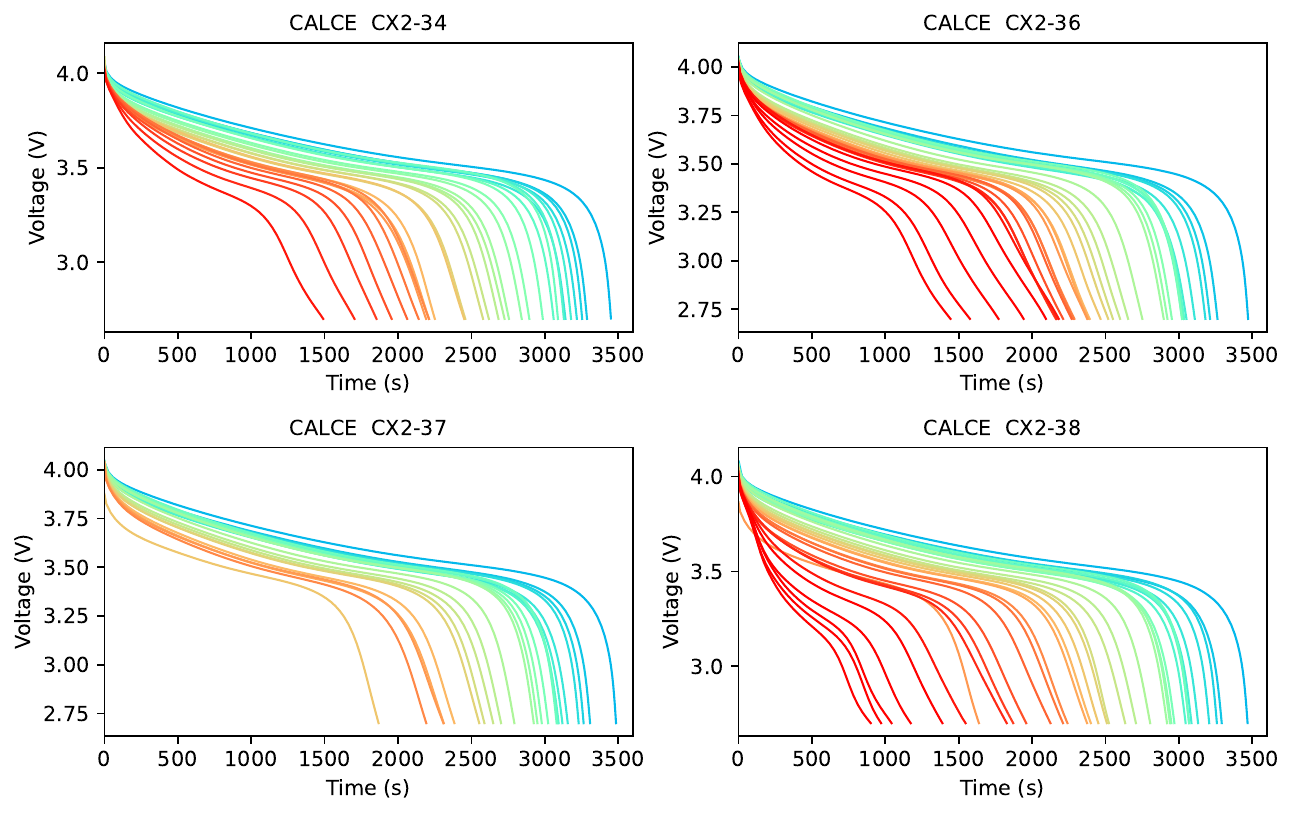} 
\caption{Discharge voltage-time (V-t) curves for the four selected CX2 batteries. A red-to-blue gradient color scheme is used to distinguish V-t curves across different cycles (CALCE CX2-34: 50--1726 cycles, CX2-36: 53--1958 cycles, CX2-37: 53--1274 cycles, CX2-38: 53--1949 cycles) during the battery's degradation process, with the progression in cycles indicated by a change in color from blue to green to orange and finally red.}
\label{fig:pineapple-dataset}
\end{figure}

In prior work, four CX2 battery cells were charged and discharged for multiple cycles using the Arbin Battery Tester and the measurements were recorded and subsequently collated in the open-source CALCE repository\footnote{CALCE battery research group homepage. URL https://calce.umd.edu/battery-data}. Each of the four cells were cycled according to a standard constant current/constant voltage (CC-CV) protocol. Specifically, a constant current rate of $0.5C$ was applied until the voltage reached $4.2V$, after which the voltage was held at $4.2V$ until the charging current decreased below $0.05A$. The batteries were then discharged at $1.35A$ until a discharge cut-off voltage of $2.7V$ was reached. 

The present study utilized the same four CX2 battery test data from the CALCE dataset, namely CX2-34, CX2-36, CX2-37, and CX2-38. The discharge voltage-time (V-t) curves for each of the four CX2 batteries are shown in Figure~\ref{fig:pineapple-dataset}. The pattern of discharge, as exemplified by the individual V-t curves in Figure~\ref{fig:pineapple-dataset}, clearly evolves as the battery ages. In the initial stage, the V-t curve exhibits a well-defined, long-lasting voltage plateau during discharge. As the battery ages, the voltage plateau shifts leftward, indicating the batteries' ability to sustain a discharge during operation has significantly reduced (i.e. capacity fade). Near the end of life, the voltage drops much more rapidly, and the curve becomes irregular or highly sloped, reflecting significant degradation.
%The data were processed to filter out overlapping and incomplete observations.

\section{First-principle model of a lithium-ion battery cell} \label{sec:first-principle}

\subsection{Single-Particle Model (SPM)}

The SPM is a first-principle description of electrochemical processes during battery charging and discharging. It assumes the lithium-ion diffusion within a battery's active material in the anode or cathode can be well-captured by the dynamics across a representative spherical particle. This model is widely used to analyze and simulate battery performance, particularly in relation to charge and discharge rates and observed voltage profiles. %, and thermal behavior

Under the SPM assumption, the transport of lithium ions inside the anode or cathode is described by Fick's Law of Diffusion. In the context of a spherical particle, it takes the following form~\cite{munteanu2024single}:
\begin{subequations} \label{eq:spm}
\small
    \begin{align}
        & \text{PDE:} & \frac{\partial\ c_k(r,t)}{\partial t} = \frac{D_k}{r^2} \ \frac{\partial}{\partial r} \left( r^2 \frac{\partial\ c_k(r,t)}{\partial r} \right), &\quad\quad r\in[0, R_k], t\in(0,T] \label{eq:spm_pde} \\
        & \text{IC:} & c_k(r,t=0) = C_k, &\quad\quad r\in[0, R_k] \label{eq:spm_ic} \\
        & \text{BC:} & \frac{\partial\ c_k(r,t)}{\partial r}\Bigr|_{r=0} = 0,\quad D_k\ \frac{\partial\ c_k(r,t)}{\partial r}\Bigr|_{r=R_k} = J_k, &\quad\quad t\in(0,T] \label{eq:spm_bc}
    \end{align}
\end{subequations} 
where $c_k(r,t)$ is the concentration of lithium ions at radial position $r$ (distance from the center of the spherical particle) within the particle and time $t$ ; $D_k$ is the diffusion coefficient of lithium ions in the cathode or anode material; $R_k$ is the particle radius. $C_k$ is the initial concentration of Li-ions in each particle at the time the discharge starts, and is commonly assumed to be spatially homogeneous.

The Neumann BCs in Equation~\ref{eq:spm_bc} enforces zero flux of lithium ions at the center of the particle ($r=0$), and a flux of magnitude $J_k$, also termed the surface current density, which represents the rate at which lithium ions diffuse across the surface of the particle at $r=R_k$. 

To simplify the notation, we use the subscript $k=p$ for the positive electrode and $k=n$ for the negative electrode. For instance, the cathode is a positive electrode (hence represented as $p$), and the anode is a negative electrode (represented as $n$) during discharge.

The surface current density of the positive electrode ($k=p$) and negative electrode ($k=n$) in Equation~\ref{eq:spm} can be represented as~\cite{munteanu2024single}:
\begin{equation} \label{eq:scd}
    J_p = \frac{I}{Fa_pA_pL_p} = \frac{I}{F} \cdot G_{p}, \quad\quad J_n = \frac{-I}{Fa_nA_nL_n} = -\frac{I}{F} \cdot G_{n}
\end{equation}
where $I$ is the applied current, $F$ is Faraday’s constant, $A_k$ the electrode area, $L_k$ the electrode thickness, and $a_k$ the specific interfacial surface area, which is given by $a_k = \frac{3\epsilon_k}{R_k}$, where $\epsilon_k$ is the volume fraction of active material in the electrodes. We further group all these information into a composite coefficient $G_k = R_k / (3\epsilon_kA_kL_k)$, given that there is no means to decouple the inference of these parameters from the macro-scale V-t measurements. \\

\subsection{Terminal voltage}

By solving Equation~\ref{eq:spm_pde},~\ref{eq:spm_ic},~\ref{eq:spm_bc} and Equation~\ref{eq:scd}, the concentration of lithium ions on the respective particle surfaces $c_k(r=R_k, t)$ can be obtained. The terminal voltage $V(t)$ of the cell can then be calculated as a function of the surface concentrations in both the positive and negative electrodes~\cite{munteanu2024single}:
\begin{multline} \label{eq:vt}
    V(t) = U_p\left(\frac{c_p(r=R_p,t)}{c_{max,p}}\right) - U_n\left(\frac{c_n(r=R_n,t)}{c_{max,n}}\right) \\
    - \frac{2R_gK}{F}\ \sinh^{-1}\left[ \frac{IG_p}{2j_p}\right] - \frac{2R_gK}{F}\ \sinh^{-1}\left[ \frac{IG_n}{2j_n}\right] + R_fI
\end{multline} 
where $K$ is the cell temperature, $R_g$ the universal gas constant, $R_f$ the internal ohmic resistance, and $j_k$ the exchange current density.

The first two terms ($U_p$ and $U_n$) in the right-side of Equation~\ref{eq:vt} are the so-called \textit{equilibrium potentials}. They describe the battery-specific evolution of normalized lithium-ion concentrations, i.e., functions of the surface concentrations with respect to battery-specific maximums ($c_{max,p}, c_{max,n}$), and have been empirically measured for different electrode materials. Their difference is also commonly referred to as the Open-Circuit Voltage (OCV): $V_{ocv}(t) = U_p(c_p(r=R_p,t)\ / c_{max,p}) - U_n(c_n(r=R_n,t)\ / c_{max,n})$. Based on the CX2 electrode materials, the $U_p$ and $U_n$ functions from ~\cite{jiang2022user} and ~\cite{gasper2022machine} were used in this work. 

The next two terms in the right-side of Equation~\ref{eq:vt} denote the positive and negative electrode \textit{over-potentials}. They are seen as an aggregated effect of resistance. Together with the lumped internal resistance, their sum is referred to as the over-voltage present in the battery, $V_{ovm}$. While ~\cite{munteanu2024single} expresses the exchange current density $j_k$ term in the positive and negative electrode \textit{over-potentials} as a function of the respective surface concentrations, the relative change in magnitudes of these terms relative to the change in OCV over each discharge cycle is small, hence they were simplified as a constant for this work. \\

\section{Physics-informed neuroevolution for prognostic parameter inference}

In this work, we propose a physics-informed neuroevolution algorithm for characterizing the battery's internal state during operation, dubbed PINEAPPLE.

\subsection{Characterizing the battery's internal state}

The calculation of terminal voltage during each discharge cycle requires the specification of several parameters (e.g. diffusivity $D_k$ and maximum lithium-ion concentration $c_{max,p}$) that characterize the internal state of the battery during operation. Inferring internal state changes that occur during operation provides valuable insights into the underlying physics of battery degradation and the state-of-health of the battery. This information both provides for a deeper understanding of battery behavior, and is a key enabler for more effective prognosis and prediction of remaining useful life.

Inferring a battery’s internal state from measurements, such as the V-t curves, presents several challenges. First, publicly available datasets often lack comprehensive information on precise electrode properties which are critical for accurate first-principle modeling. Additionally, experimental test data typically includes measurement noise, missing data, and inconsistencies due to differences in testing conditions or protocols or simply inter-battery variability, making it difficult to directly apply mechanistic models without extensive data processing and calibration, even across batteries of the same type. These limitations make it challenging to infer battery internal states from real-world data without careful calibration and model adaptation to ensure meaningful insights. %key parameters, such as the anode and cathode materials, or exchange current densities,
%Terminal voltage calculations further rely on empirically derived OCV and over-voltage functions, but they usually do not generalize well across different battery materials and fail to evolve with the degradation process.

To calibrate the first-principle model to the measured V-t curves, we introduce \textit{scaling factors} and apply them to several parameters critical to the SPM and terminal voltage calculation, including the diffusion coefficient, geometrical composite coefficient, maximum lithium-ion concentration, initial lithium-ion concentration, particle radius, and exchange current density. This organization is consistent with prior work suggesting that the dynamics in each particle of the SPM can be reduced to 2 unique parameters after appropriate non-dimensionalization ~\cite{mendez2024physics}. These scaling factors serve to compensate for real-world, inter-battery variability and provide a flexible means for generalizing the framework to provide cycle-dependent insights across different batteries without precise electrode characterization. %For example, we denote the exchange current density $j_k=\eta_{j_k} \tilde{j}_k$, where $\eta_{j_k}$ is a scaling factor and $\tilde{j}_k$ is a reference value obtained from literature. 

These parameters can be empirically estimated at each cycle, and the evolution of these parameters from the inferred value obtained during the initial (or early) cycle(s) for each battery serves as a means of understanding the impact of charge/discharge cycles and any cumulative degradation; details are summarized in Table~\ref{tab:pineapple-params}. For this work, four cycle-dependent parameters are identified based on an initial empirical sensitivity analysis, namely the diffusion coefficients for both the positive and negative electrodes ($D_p=\eta_{D_p} \tilde{D}_p$, $D_n=\eta_{D_n} \tilde{D}_n$), and the geometrical composite coefficient ($G_p=\eta_{G_p} \tilde{G}_p$) and maximum lithium-ion concentration ($c_{max,p}=\eta_{c_{max,p}} \tilde{c}_{max,p}$) for the positive electrode. To account for their variation during the degradation process, we define a \textit{hypothetical range} of scaling values rather than fixed values for these four cycle-dependent parameters based on their respective physics. For example, the geometrical composite coefficient cannot undergo orders of magnitude variation in the size of the particle, whereas the diffusion coefficients reported in literature commonly span several orders of magnitude. Consequently, \textit{the problem of characterizing the batteries' internal state during operation (and hence degradation) reduces to inferring these four scaling factors:} $\eta_{D_p}$, $\eta_{D_n}$, $\eta_{G_p}$, and $\eta_{c_{max,p}}$, for V-t curves obtained during the battery's lifetime.

\begin{figure}[!hbt]
\centering
\includegraphics[width=1.\linewidth]{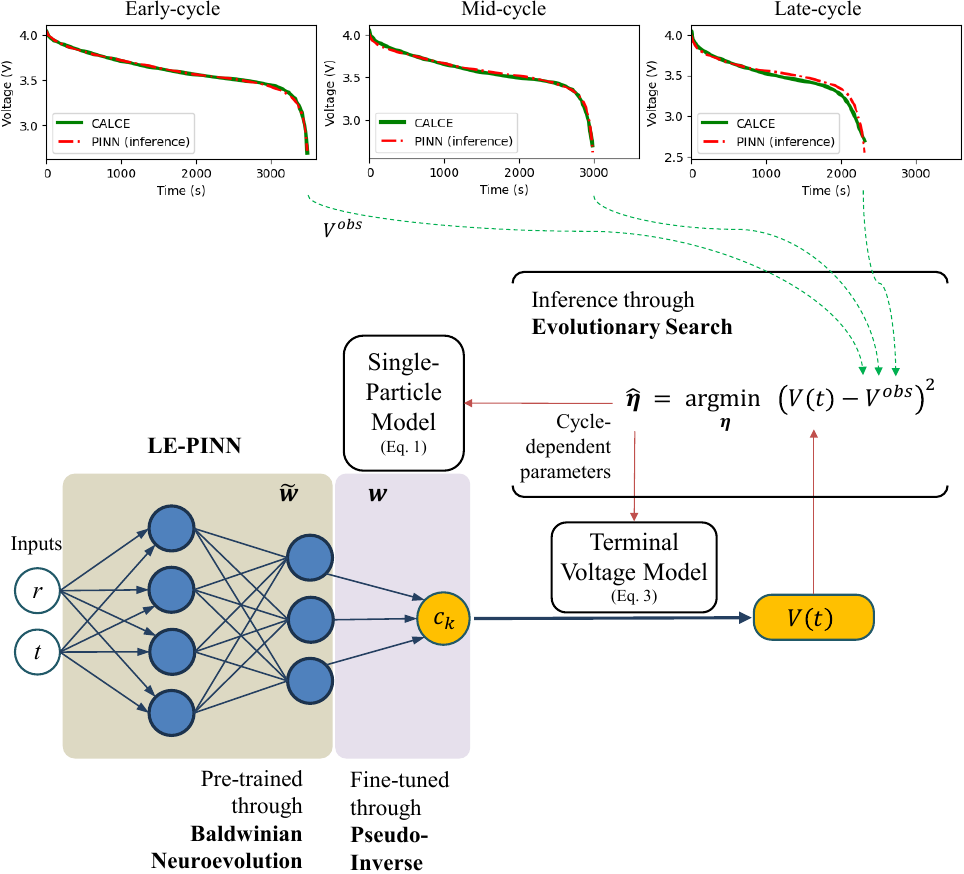} 
\caption{PINEAPPLE schematic. The nonlinear hidden layers of LE-PINN are meta-learned (pre-trained) through Baldwinian neuroevolution, which takes less than 600 seconds on a sparse simulation dataset with diffusion coefficients ($D_k$) spanning 3 orders of magnitude. The fine-tuning (forward solve) of the final layer for positive and negative electrodes together costs less than 20 milliseconds for specific SPM parameters. The evolutionary search for best fit cycle-dependent parameters to match the observed V-t curve, $V^{obs}$, can be completed in a few seconds. Representative examples for early-, mid-, and late-cycle experimentally observed V-t curves (obtained from CALCE CX2-37 Cycle 53, Cycle 638, and Cycle 1181 respectively) and the corresponding prediction from the PINEAPPLE framework are illustrated in the top portion of the Figure.}
\label{fig:pineapple-schemetic}
\end{figure}

An instantiation of PINEAPPLE is demonstrated through the inference of four cycle-dependent scaling factors $\eta_{D_p}$, $\eta_{D_n}$, $\eta_{G_p}$, and $\eta_{c_{max,p}}$ from measured V-t curves, although more parameters can be flexibly included. A schematic of PINEAPPLE is presented in Figure~\ref{fig:pineapple-schemetic}, comprising two main stages: an offline meta-learning phase and an online inference phase.

%%%%%%%%%%%%%%%%%%%%%%%%%%%%%%%%%%%%%%%%%%%%%%%%%%%% FIG %%%%%%%%%%%%%%%%%%%%%%%%%%%%%%%%%%%%%%%%%%%%%%%%%%%%

\subsection{PINEAPPLE-pretrain: offline meta-learning of a generalizable PINN for lithium-ion batteries}

The first component of the PINEAPPLE framework is the training of a generalizable Lithium-ion battery Electrochemical Physics-Informed Neural Network (LE-PINN) model for predicting lithium-ion concentration $c_k(r,t)$ across varying SPM parameters, including different PDE and BC combinations for the positive and negative electrodes. 

While conventional PINNs can manifest slow convergence and limited generalization to unseen physics conditions, we utilize a Baldwinian-PINN framework in this work~\cite{wong2023generalizable}. This meta-learning PINN approach is versatile and can be rapidly fine-tuned to provide zero-shot and fast predictions for new conditions (e.g., over a range of cycle-dependent SPM parameters) with high accuracy. Crucially, it was previously shown that Baldwinian-PINNs are fast, data-efficient, and provide more versatile and accurate predictions for new tasks relative to other meta-learning PINNs, making them the model of choice in PINEAPPLE.

As an algorithmic realization of Baldwinian evolution in the context of meta-learning PINNs, PINEAPPLE-pretrain features an \textit{outer evolutionary optimization} loop, where a population of LE-PINN models is collectively exposed to physics conditions sampled from a probability distribution over relevant conditions (i.e. specific SPM parameters). Models that excel in a randomly selected subset of these conditions during the \textit{inner physics-informed fine-tuning} process are deemed fitter for survival, creating selection pressure toward network weights that reinforce desirable learning biases. We further describe these two key components, (i) \textit{inner physics-informed fine-tuning}, and (ii) \textit{outer evolutionary optimization}, below.

\textbf{Physics-informed fine-tuning.} Without loss of architectural generality, we can write the LE-PINN model's final output $c(r,t)$ as:
\begin{equation}
    c(r,t) = \sum_j w_j f_j(r,t;\boldsymbol{\tilde{w}})
\end{equation}
where $\boldsymbol{w} = [\dots w_j \dots]^T$ are the linear output layer weights. Here, $f_j(r,t;\boldsymbol{\tilde{w}})$ denotes the $j=1,2,...,n_{\text{hidden nodes}}$ nonlinear hidden layers (prior to final output layer), and $\boldsymbol{\tilde{w}}$ denotes the network weights from the nonlinear hidden layers collectively. %with proven representation capacity
%LE-PINN's nonlinear hidden layers (prior to final output layer) as $f_j(r,t;\boldsymbol{\tilde{w}})$, where $j=1,2,...,n_{\text{hidden nodes}}$ with proven representation capacity, and let $\boldsymbol{\tilde{w}}$ denotes the network weights from the nonlinear hidden layers collectively. The final output $c(r,t)$ can be written as:
As per previous work, we limit the fine-tuning process to the final layer, i.e., finding the best set of $\boldsymbol{w}$ such that the LE-PINN’s outputs satisfy SPM for a specific physics condition (i.e. specific set of SPM parameters). Since the SPM governing equations are all linear with respect to $c(r,t)$, it is possible to utilize a Tikhonov regularization procedure (Pseudo-Inverse) for achieving fast (order of milli-seconds in our experiments), accurate, and physics-compliant solutions by solving the following least-squares problem:
%\begin{small}
\begin{align}
    \left[
    \begin{array}{ccc}
        \dots & \lambda_{\text{PDE}}\ \frac{\partial\ f_j(r,t;\boldsymbol{\tilde{w}})}{\partial t} - \frac{D_k}{r^2} \ \frac{\partial}{\partial r} \left( r^2 \frac{\partial\ f_j(r,t;\boldsymbol{\tilde{w}})}{\partial r} \right) & \dots \\ 
        \vdots & \vdots & \vdots   \\
        \dots & \lambda_{\text{IC}}\ f_j(x,0;\boldsymbol{\tilde{w}}) & \dots \\ 
        \vdots & \vdots & \vdots   \\     
        \dots & \lambda_{\text{BC}}\ \frac{\partial\ f_j(0,t;\boldsymbol{\tilde{w}})}{\partial t} & \dots \\ 
        \vdots & \vdots & \vdots   \\
        \dots & \lambda_{\text{BC}}\ D_k\ \frac{\partial\ f_j(R_k,t;\boldsymbol{\tilde{w}})}{\partial t} & \dots \\
    \end{array}
    \right] 
    \left[
    \begin{array}{c}
        \vdots \\
        w_j    \\
        \vdots \\
    \end{array}
    \right]    
    &=
    \left[
    \begin{array}{c}
        0  \\ 
        \vdots  \\ 
        \lambda_{\text{IC}}\ C_k  \\ 
        \vdots  \\ 
        0  \\ 
        \vdots  \\ 
        \lambda_{\text{BC}}\ J_k  \\ 
    \end{array}
    \right] \nonumber \\
    \boldsymbol{A} \boldsymbol{w} &= \boldsymbol{b} \label{eq:leastsquares} 
\end{align}
%\end{small}

The matrix $\boldsymbol{A}$ and vector $\boldsymbol{b}$ are a weighted system of equations constructed by substituting $f_j(r,t;\boldsymbol{\tilde{w}})$ into the SPM governing equations (Equation~\ref{eq:spm}), for a set of collocation points $(r,t)$. Moreover, $\lambda_{\text{PDE}}$, $\lambda_{\text{IC}}$, and $\lambda_{\text{BC}}$ are hyper-parameters empirically selected to adjust the relative weighting of PDE, IC, and BC constraints. Note that the derivatives required to construct matrix $\boldsymbol{A}$ can be easily computed by automatic differentiation. Such a least-squares formulation yields a closed-form result in a single computation step:
\begin{subequations} \label{eq:pseudoinverse}
    \begin{align}
        \boldsymbol{w} &= (\lambda_{\text{PI}} I + \boldsymbol{A}^T\boldsymbol{A})^{-1} \boldsymbol{A}\boldsymbol{b} &,  \text{if $\boldsymbol{A}$ is over-determined} \\
        \boldsymbol{w} &= \boldsymbol{A}^T (\lambda_{\text{PI}} I + \boldsymbol{A}\boldsymbol{A}^T)^{-1} \boldsymbol{b} &, \text{if $\boldsymbol{A}$ is under-determined}
    \end{align}
\end{subequations} 
The learning hyperparameter $\lambda_{\text{PI}} \geq 0$ reduces the $L^2$-norm of the least-squares solution. The inclusion of the Tikhonov regularization, $\lambda_{\text{PI}} \geq 0$, can significantly improve the conditioning of the problem, thereby improving the quality of the obtained solution. Additionally, Equation~\ref{eq:pseudoinverse} reduces the matrix size for inversion, further accelerating the time to solution.

After determining the solution for $\boldsymbol{w}$, the concentration values can be obtained by performing a forward pass to the network $c(r,t;\boldsymbol{\tilde{w}},\boldsymbol{w})$. This fine-tuning process is purely physics-based (requiring no ground truth labels) and is exceptionally fast when performed for each new set of SPM parameters.

\textbf{Outer evolutionary optimization.} The goal of the outer evolutionary optimization is to obtain an optimal weights distribution $\boldsymbol{\tilde{w}}$ which serves as the foundation for generating accurate fine-tuned models, as measured by both predictive performance and physics-consistency with respect to SPM, over a diverse set of tasks (varying SPM parameters). We thus define the outer loop optimization objective as below:
\begin{equation}
    \mathcal{L}(\boldsymbol{\tilde{w}}) = \sum_{\text{subset task}} \left( \tau_{LSE} \ \mathcal{L}_{LSE}(\boldsymbol{\tilde{w}}) + \tau_{MSE} \ \mathcal{L}_{MSE}(\boldsymbol{\tilde{w}}) \right) \label{eq:ea-loss}
\end{equation}
where $\mathcal{L}(\boldsymbol{\tilde{w}})$ is the weighted ($\tau_{LSE}$, $\tau_{MSE}$) sum of the least-squares error (LSE), 
\begin{equation}
    \mathcal{L}_{\text{LSE}}(\boldsymbol{\tilde{w}}) = \left(\boldsymbol{A}(\boldsymbol{\tilde{w}}) \boldsymbol{w} - \boldsymbol{b}\right)^T \left(\boldsymbol{A}(\boldsymbol{\tilde{w}}) \boldsymbol{w} - \boldsymbol{b}\right)
\end{equation}
that indicates the degree of physics compliance, and the mean squared error (MSE) between LE-PINN output $c(r,t;\boldsymbol{\tilde{w}},\boldsymbol{w})$ and training label $c^{label}$ over a set of collocation points $({r,t})$,
\begin{equation}
    \mathcal{L}_{\text{MSE}}(\boldsymbol{\tilde{w}}) = \sum_{r, t} \left( c(r,t;\boldsymbol{\tilde{w}},\boldsymbol{w}) - c^{label} \right)^2
\end{equation}
During the optimization procedure, these errors are aggregated over a randomly selected subset of training tasks to indicate model performance. Furthermore, the outer evolutionary optimization can be extended to determine optimal learning and weight balancing hyperparameters $\boldsymbol{\lambda}=$($\lambda_{\text{PI}}$, $\lambda_{\text{PDE}}$, $\lambda_{\text{IC}}$, $\lambda_{\text{BC}}$) as required for the physics-informed fine-tuning. %to indicate the predictive performance.

Baldwinian neuroevolution yields pre-trained weights for the nonlinear hidden layer and learning hyperparameters ($\boldsymbol{\tilde{w}}$, $\boldsymbol{\lambda}$) from the training tasks, enabling LE-PINN to deliver fast and accurate zero-shot predictions for new set of SPM parameters.

\textbf{Implementation details.} Our implementation of the LE-PINN architecture follows the approach proposed in ~\cite{wong2023generalizable}. The model employs a shallow network with a single hidden layer utilizing a combination of smooth nonlinear activation functions, including $sin$, $silu$, and $tanh$. The weight distribution $\boldsymbol{\tilde{w}}$ is drawn from a probability distribution governed by a dimensionally reduced set of network hyperparameters $\boldsymbol{\theta}$ which define the mean and spread of weights and biases across different network blocks. By optimizing distribution parameters for weight groups rather than individual weights in the nonlinear hidden layer, the search space is significantly reduced, enabling highly efficient exploration using state-of-the-art evolutionary algorithms. We employ the CR-FM-NES optimizer implemented with a JAX backend to leverage performance improvements in automatic differentiation and linear algebra operations \cite{nomura2022fast,bradbury2018jax,evojax2022}.

The weighting coefficients for the loss terms in outer loop optimization objective (Equation~\ref{eq:ea-loss}) are fixed throughout the evolutionary process. The outer-loop loss weights are set to $\tau_{LSE}$=0.1 and $\tau_{MSE}$=1.0, and are not adaptively adjusted during training. These values were selected through light empirical tuning to balance the relative magnitudes of the physics-based loss (LSE) and the data-driven MSE. In our experiments, this fixed weighting scheme was sufficient for stable convergence and satisfactory accuracy across the tested tasks, although proper scaling between different loss components has been reported in the PINN literature to significantly impact convergence behavior and training stability. The meta-learning process is conducted on a workstation equipped with 4 $\times$ NVIDIA L40 GPUs.

\subsection{PINEAPPLE-inference: online inference of battery's internal state parameters}

The objective at inference is to estimate variations in the four scaling factors $\eta_{D_p}$, $\eta_{D_n}$, $\eta_{G_p}$, and $\eta_{c_{max,k}}$ which correspond to key internal state parameters $D_p$, $D_n$, $G_p$, and $c_{max,k}$ as per the SPM and terminal voltage function encapsulated in Equation~\ref{eq:vt}. This estimation is formulated as a search problem, where the goal is to find optimal scaling factors that best minimize the squared distance (best-fit) between the measured V-t curve, $V^{\text{obs}}$, and the predicted V-t curve, $V(t)$, as generated by the LE-PINN:

\begin{equation}
    \hat{\eta}_{D_p},\hat{\eta}_{D_n},\hat{\eta}_{G_p},\hat{\eta}_{c_{max,k}} = \argmin_{\eta_{D_p},\eta_{D_n},\eta_{G_p},\eta_{c_{max,k}}} \left( V(t) - V^{\text{obs}}  \right)^2
\end{equation}

Crucially, for each candidate solution comprising the four scaling factors, the SPM provides a fundamental, mechanistic prediction of how $\eta_{D_p}$, $\eta_{D_n}$, and $\eta_{G_p}$ impact surface lithium-ion concentrations across both positive and negative electrodes. Through incorporation of this physics in the fine-tuning process, the pre-trained LE-PINN can rapidly infer the surface lithium-ion concentrations with high accuracy, eliminating the need for more computationally expensive simulations while ensuring that the predicted battery response remains physically consistent with the underlying electrochemical dynamics. Once the surface lithium-ion concentrations are obtained, $\eta_{G_p}$ and $\eta_{c_{max,k}}$ are incorporated into the terminal voltage model to compute the corresponding V-t curve. An advantage of our methodology is that LE-PINN predictions can be flexibly evaluated at irregularly sampled locations to align with measurement points as required, negating the need for any further interpolation or subsampling of measurement points.

To effectively solve this optimization problem through evolutionary search, we employ the state-of-the-art Covariance Matrix Adaptation Evolution Strategy (CMA-ES) algorithm~\cite{hansen2016cma}, known for its robust performance in high-dimensional, non-convex optimization tasks. CMA-ES iteratively refines candidate solutions based on covariance matrix adaptation, ensuring efficient exploration and convergence toward optimal scaling factors. Given the exceptionally fast adaptation capabilities of the LE-PINN model and the efficient parameter space exploration enabled by CMA-ES, we are able to execute an evolutionary search over 50 generations (with a population size of 20) in under 5 seconds. This inference speed can be further accelerated through hardware-level parallelization, underscoring the potential of our proposed PINEAPPLE framework for real-time battery parameter inference and adaptive control, in addition to providing an interpretable understanding of how variations in the battery's internal state relate to degradation. In addition, the evolutionary search algorithm can be seen as an adaptation of the initialized distribution into a proposal distribution concentrated in high-likelihood regions. This distribution can be seen as an approximation to the Bayesian posterior, though it is obtained via optimization rather than exact Bayes’ rule. %---sufficient to achieve well-converged solutions--- influence its performance during its

\section{Results and discussion}

In order to evaluate the effectiveness of the PINEAPPLE framework, we apply our method for inferring the evolution of each battery's internal state to four batteries (CX2-34, CX2-36, CX2-37, CX2-38) from the public CALCE data repository. 

\subsection{SPM and terminal voltage model parameters for CX2 battery}

Before initiating the meta-learning process for the LE-PINN, it is helpful to determine an approximate parameter range for the SPM and terminal voltage model as pertaining to the real-world dataset of interest. For the CX2 battery (as presented in Section~\ref{sec:dataset}), the cathode material is Lithium Cobalt Oxide (LCO), while the anode material is Graphite. Reference values from literature for the corresponding electrode properties and electrochemical parameters, and their associated scaling factors, are summarized in Table~\ref{tab:pineapple-params}. Through a simple sensitivity analysis, we identified an expansive range for the 4 cycle-dependent scaling factors, and fixed the other parameters' scaling factors accordingly to ensure consistency with the measured data. It should be noted that the framework is fundamentally agnostic to the choice of scaling factors, although prior domain knowledge as in \cite{mendez2024physics} can help to focus the dimensionality of the inverse problem. Crucially, Baldwinian-PINNs do not require \textit{a priori} parameterizations to be defined as they can be trained through an evolutionary algorithm on any training task distribution. %the simple sensitivity analysis employed in this work presumes the availability of some initial experimental data for the battery type of interest.

\subsection{Training dataset for offline meta-learning}

In this work, the training and test datasets for offline meta-learning of the LE-PINN model comprise 150 simulation tasks with varying PDE and BC configurations for the positive and negative electrodes, including diffusion coefficients spanning three orders of magnitude. They are generated using PyBaMM\footnote{Python Battery Mathematical Modelling (PyBaMM): an open-source battery simulation package written in Python. URL https://github.com/pybamm-team/PyBaMM.}~\cite{sulzer2021python}. For each simulation task, the concentration of lithium ions is solved for a time domain between 0 and $T$=3600 seconds, at a 60 second interval (61 grid points), while the spatial resolution employed comprised 64 grid points between 0 and $R_k$. More tasks are used for training the LE-PINN model for the positive electrode as the LE-PINN model has to learn across 2 different parameters (i.e. both $D_p$, $G_p$). 

For the positive electrode, 100 simulations are generated with diffusion coefficients and composite coefficients randomly sampled from the log space $D_p \in [3.9\mathrm{e}^{-15}, 3.9\mathrm{e}^{-13}]$ and the linear space $G_p \in [1.01, 4.03]$ respectively. 50 simulations are separately generated for the negative electrode with diffusion coefficients randomly sampled from the log space $D_n \in [3.9\mathrm{e}^{-16}, 3.9\mathrm{e}^{-13}]$. These simulated tasks cover a range of scaling factors, i.e., $\eta_{D_p} \in [1\mathrm{e}^{-1} - 1\mathrm{e}^{1}]$, $\eta_{D_n} \in [1\mathrm{e}^{-2} - 1\mathrm{e}^{1}]$, $\eta_{G_p} \in [1, 4]$, for the SPM parameters of interest ($D_p$, $D_n$, $G_p$). 

We then randomly split the 150 simulation tasks into 100 (60 positive electrodes and 40 negative electrodes) training tasks and 50 (40 positive electrodes and 10 negative electrodes) test tasks. Using the 100 training tasks, the meta-learning process for the nonlinear hidden layers of LE-PINN and its learning hyperparameters can be completed in under 600 seconds.

%%%%%%%%%%%%%%%%%%%%%%%%%%%%%%%%%%%%%%%%%%%%%%%%%%%% FIG %%%%%%%%%%%%%%%%%%%%%%%%%%%%%%%%%%%%%%%%%%%%%%%%%%%%
\begin{figure}[!hbt]
\centering
\includegraphics[width=1.0\linewidth]{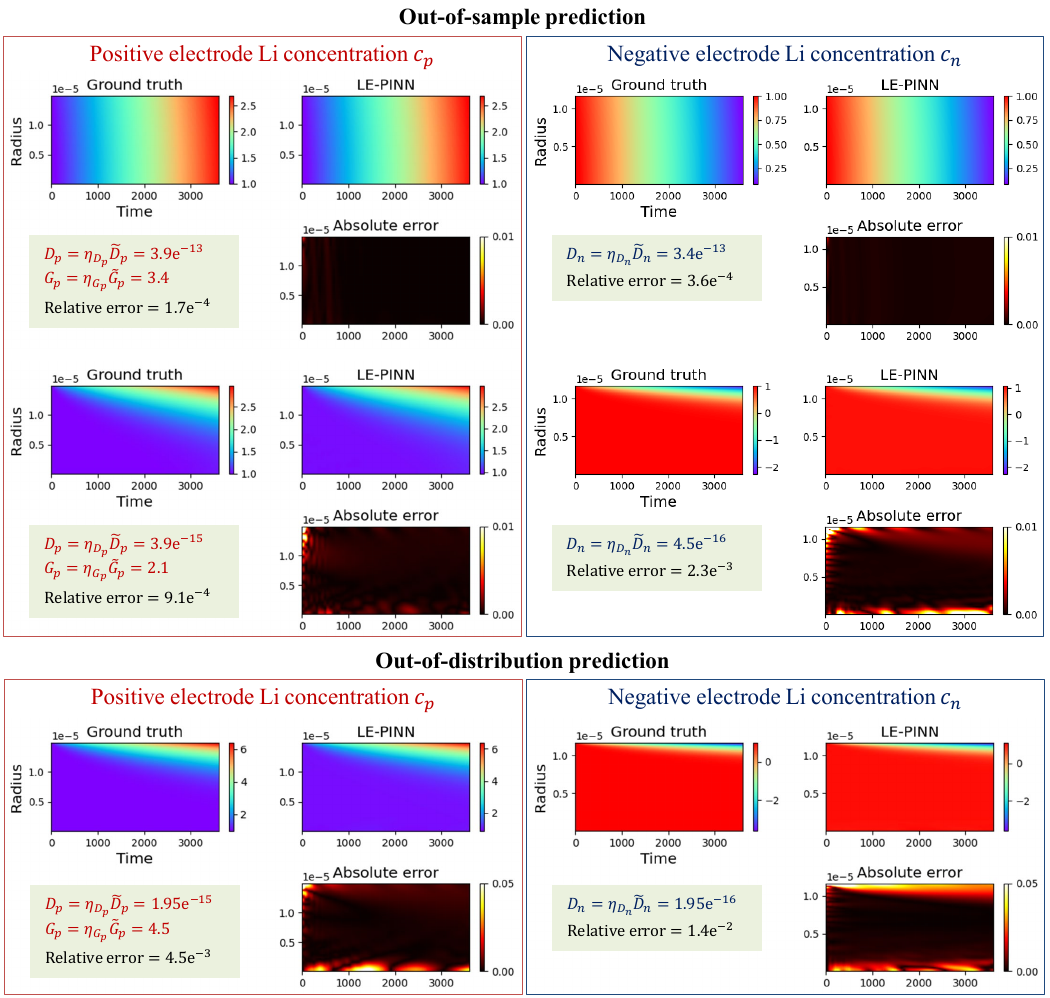} 
\caption{Predictive performance of LE-PINN model: examples of the solution vs. ground truth for unseen (top: out-of-sample, bottom: out-of-distribution) SPM parameters for positive and negative electrodes. The concentrations $C_p$ and $C_n$ are normalized by their respective initial values.}
%and (b) simulation of discharge V-t curves using different SPM and terminal voltage calculation parameters.}
\label{fig:pineapple-forward}
\end{figure}

%%%%%%%%%%%%%%%%%%%%%%%%%%%%%%%%%%%%%%%%%%%%%%%%%%%% TAB %%%%%%%%%%%%%%%%%%%%%%%%%%%%%%%%%%%%%%%%%%%%%%%%%%%%
\begin{table}[htbp]
\centering
\caption{Predictive performance of LE-PINN model.} \label{tab:pineapple-forward}
\footnotesize
\begin{tabular*}{\textwidth}{@{\extracolsep{\fill}}>{\raggedright\arraybackslash}p{2.2cm}>{\raggedright\arraybackslash}p{2.5cm}>{\raggedright\arraybackslash}p{1.5cm}>{\raggedleft\arraybackslash}p{2.2cm}>{\raggedleft\arraybackslash}p{2.2cm}>{\raggedleft\arraybackslash}p{1.5cm}}
    \toprule
    {} & SPM parameter & Scaling factor & Relative error (train) & Relative error (test) & Evaluation time$^{1}$ \\
    \cmidrule{1-6}  
    {\textbf{Positive electrode$^{2}$, $C_p$}} & $D_{p} \in [3.9\mathrm{e}^{-15}, 3.9\mathrm{e}^{-13}]$ $G_{p} \in [1.01, 4.03]$ & $\eta_{D_{p}} \in [1\mathrm{e}^{-1}, 1\mathrm{e}^{1}]$ $\eta_{G_{p}} \in [1, 4]$ & $3.57\mathrm{e}^{-4}$ {\tiny$\pm 2.99\mathrm{e}^{-4}$} & $2.78\mathrm{e}^{-4}$ {\tiny$\pm 2.61\mathrm{e}^{-4}$} & 5.62 ms {\tiny$\pm 4.43$ ms} \\
    \cmidrule{1-6}  
    {\textbf{Negative electrode$^{3}$, $C_n$}} & $D_{p} \in [3.9\mathrm{e}^{-16}, 3.9\mathrm{e}^{-13}]$ & $\eta_{D_{n}} \in [1\mathrm{e}^{-2}, 1\mathrm{e}^{1}]$ & $7.94\mathrm{e}^{-4}$ {\tiny$\pm 6.29\mathrm{e}^{-4}$} & $1.29\mathrm{e}^{-3}$ {\tiny$\pm 6.96\mathrm{e}^{-4}$} & 5.64 ms {\tiny$\pm 4.38$ ms} \\
    \bottomrule
\end{tabular*}
\raggedright
\footnotesize
$^{1}$ The evaluation time per case on single NVIDIA L40 GPU. \\
$^{2}$ $n_{\text{train}}=60$, $n_{\text{test}}=40$ positive electrode tasks. \\ 
$^{3}$ $n_{\text{train}}=40$, $n_{\text{test}}=10$ negative electrode tasks.
\end{table}

\subsection{Predicting lithium-ion concentrations of the electrodes}

As illustrated in Figure~\ref{fig:pineapple-forward}, Baldwinian evolution enables LE-PINN model to rapidly learn highly accurate solutions (i.e., lithium-ion concentrations for a given SPM parameter for a single positive or negative electrode) on new test tasks through physics-informed fine-tuning of the final layer. The predictive performance of LE-PINN model is assessed on the 50 unseen (out-of-sample) test tasks. The quantitative results are summarized in Table~\ref{tab:pineapple-forward}. As a key metric in scientific computing, we also report the relative (L2) error, given its normalization by the scale of the ground truth's norm and its consequent ease of comparison across tasks with different magnitudes. The learned solutions achieve an average relative error of 2.78$\mathrm{e}^{-4}$ {\footnotesize$\pm$2.61$\mathrm{e}^{-4}$} for the 40 positive electrode tasks and 1.29$\mathrm{e}^{-3}$ {\footnotesize$\pm$6.96$\mathrm{e}^{-4}$} for the 10 negative electrode tasks. The test errors are of a similar magnitude to the training errors, demonstrating the effectiveness of the approach. 

%%%%%%%%%%%%%%%%%%%%%%%%%%%%%%%%%%%%%%%%%%%%%%%%%%%% FIG %%%%%%%%%%%%%%%%%%%%%%%%%%%%%%%%%%%%%%%%%%%%%%%%%%%%
\begin{figure}[!hbt]
\centering
\includegraphics[width=1.\linewidth]{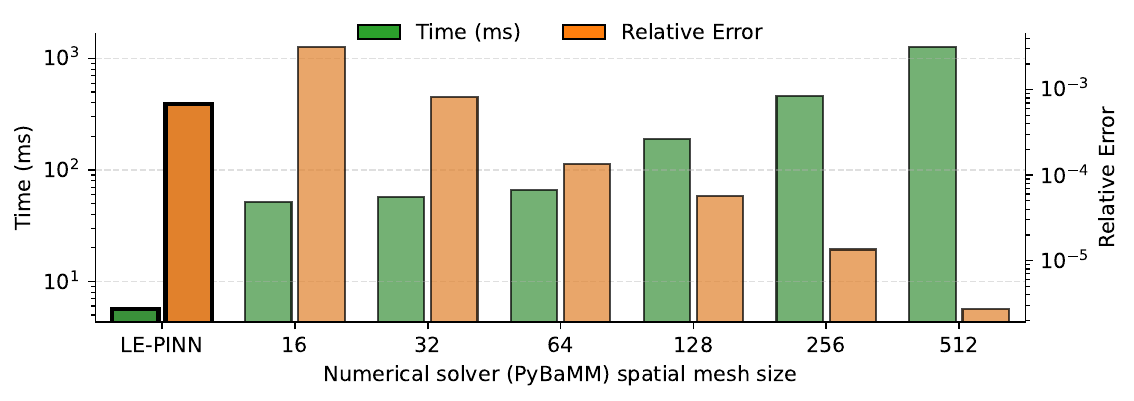} 
\caption{Comparison of relative error and computational time between LE-PINN (obtained via fine-tuning of the final layer) and numerical solver from PyBaMM at different mesh resolutions, using an out-of-sample SPM parameter for the positive electrode as a representative example. LE-PINN achieves comparable accuracy to that of the numerical solution with a mesh size of 32, while requiring significantly less (10$\times$) computational time. For all cases, errors are computed with respect to the reference numerical solution generated by PyBaMM using a mesh size of 1024.}
\label{fig:pineapple-benchmark-time-error}
\end{figure}

Predicting lithium-ion concentration for a new SPM parameter using LE-PINN (which involves fine-tuning the final layer) requires less than 6 milliseconds (ms) on a single NVIDIA L40 GPU. Figure~\ref{fig:pineapple-benchmark-time-error} compares the solution accuracy and computational time of LE-PINN with those of numerical solutions obtained using PyBaMM across a range of mesh resolutions. In this work, the PyBaMM scipy solver is used with relative and absolute tolerances set to $\text{rtol} = 1\mathrm{e}^{-6}$ and $\text{atol} = 1\mathrm{e}^{-8}$. The computational time for LE-PINN is 5.62 ± 4.42 ms, aggregated over 25 repeated runs. The out-of-sample prediction accuracy of LE-PINN is comparable to that of the numerical solution computed with a mesh size of 32, which requires 57.1 ± 13.8 ms (aggregated over 25 repeated runs) when executed on an Intel Xeon Gold 6238R @ 2.20 GHz CPU. This order-of-magnitude (10$\times$) reduction in computational time at comparable accuracies highlights the efficiency of the LE-PINN approach.

To further assess the robustness of LE-PINN beyond the parameter distributions used during meta-learning, we conduct an out-of-distribution (OOD) generalization study, with results included in Figure~\ref{fig:pineapple-forward} for both the positive and negative electrodes. For the positive electrode, we consider an OOD test case with $D_p=1.95\mathrm{e}^{-15}$ and $G_p=4.5$, which lies outside the training ranges of $D_p \in [3.9\mathrm{e}^{-15}, 3.9\mathrm{e}^{-13}]$ and $G_p \in [1.01, 4.03]$. This choice is physically motivated, as battery degradation is commonly associated with reduced diffusion coefficient $D_p$ and increased geometrical composite coefficient $G_p$. Similarly, for the negative electrode, we evaluate an OOD case with $D_n=1.95\mathrm{e}^{-16}$, which is outside the training distribution for $D_n \in [3.9\mathrm{e}^{-16}, 3.9\mathrm{e}^{-13}]$.

As expected, the relative errors for these 2 cases are higher than those observed for the in-distribution, out-of-sample test tasks. Nevertheless, LE-PINN remains stable and produces accurate solutions after physics-informed fine-tuning of the final layer, i.e., relative error of 4.5$\mathrm{e}^{-3}$ and 1.4$\mathrm{e}^{-2}$ for positive and negative electrode cases, respectively. Notably, for both electrodes, the prediction accuracy remains comparable to that of the numerical solution obtained using PyBaMM with a mesh size of 32, consistent with the performance observed in the out-of-sample test cases. These results suggest that while the zero-shot performance of LE-PINN could naturally degrade outside the training distribution, the meta-learned representation retains sufficient flexibility to generalize to physically meaningful out-of-distribution scenarios through rapid, zero-shot physics-informed adaptation. %(the trend reflecting increased difficulty of the underlying physics under more extreme parameter regimes)

%%%%%%%%%%%%%%%%%%%%%%%%%%%%%%%%%%%%%%%%%%%%%%%%%%%% FIG %%%%%%%%%%%%%%%%%%%%%%%%%%%%%%%%%%%%%%%%%%%%%%%%%%%%
\begin{figure}[!hbt]
\centering
\includegraphics[width=1.0\linewidth]{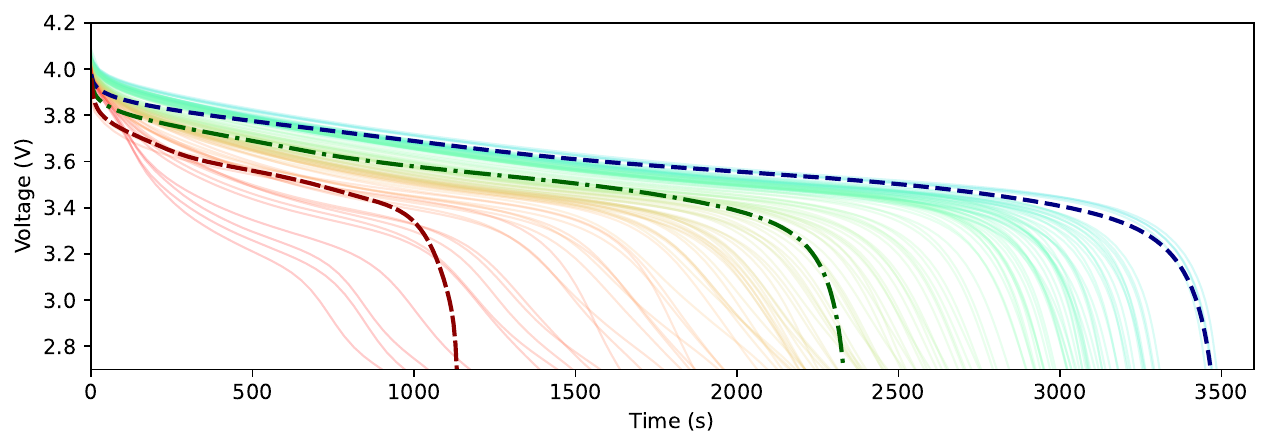} 
\caption{Simulated discharge V-t curves (dashed) using different SPM and terminal voltage calculation parameters. These curves are used for a preliminary evaluation before conducting inverse inference on the real-world dataset. The simulated curves correspond well qualitatively to the early (blue), middle (green), and late (red) cycles of the battery degradation process. The range of measured V-t curves, from the real dataset as described in Section~\ref{sec:dataset}, is shown in the background as semi-transparent solid lines.}
\label{fig:pineapple-validate-data}
\end{figure}

%%%%%%%%%%%%%%%%%%%%%%%%%%%%%%%%%%%%%%%%%%%%%%%%%%%% FIG %%%%%%%%%%%%%%%%%%%%%%%%%%%%%%%%%%%%%%%%%%%%%%%%%%%%
\begin{figure}[!hbt]
\centering
\includegraphics[width=1.0\linewidth]{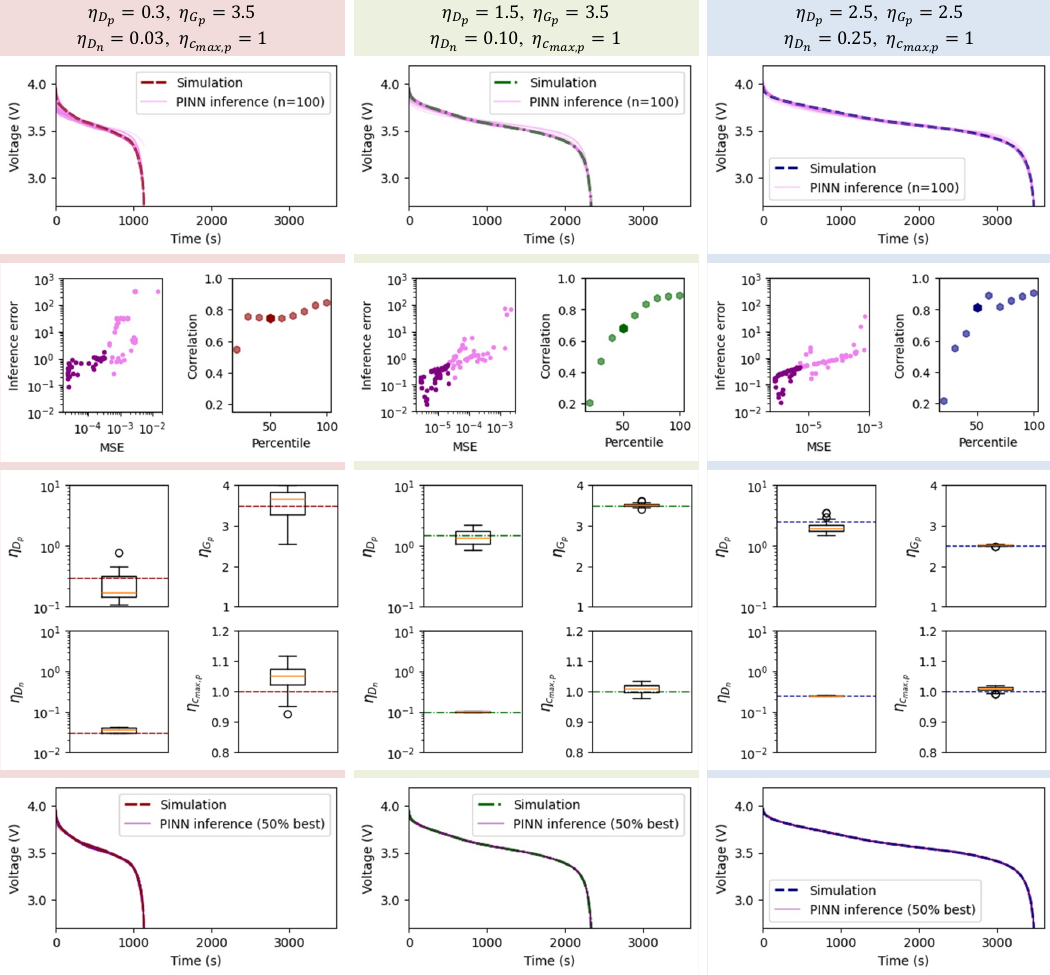} 
\caption{Validation of the PINEAPPLE framework on three synthetic V-t curves, representing the early (blue), middle (green), and late (red) stages of the battery degradation process. The first row of plots compares the synthetic V-t curves with LE-PINN-predicted V-t curves, based on inferred parameters obtained from 100 optimization runs with different random initializations. The second row illustrates the correlation between inference error (i.e., deviation of inferred scaling factors from ground truth) and MSE (difference between synthetic and LE-PINN-predicted V-t curves). Data points with MSE values below the 50th percentile threshold are highlighted in dark purple. Additionally, Spearman’s rank correlation coefficients for varying MSE percentile thresholds are provided. The third row presents the distribution of inferred scaling factors, excluding the worst 50$\%$ of MSE cases (which suggest poorer convergence during the evolutionary search). The dotted lines indicate their respective ground truth values. The final row overlays the synthetic V-t curves with the LE-PINN-predicted V-t curves after filtering out the highest 50$\%$ MSE cases.}
\label{fig:pineapple-validation}
\end{figure}

\subsection{Estimating cycle-dependent scaling factors from V-t curves}

\subsubsection{Simulated V-t curves.} Before performing inverse inference on the CALCE dataset, we first validate the PINEAPPLE framework using synthetic  V-t curves. These V-t curves are generated by solving the SPM with PyBaMM for both positive and negative electrodes, followed by passing the resulting surface lithium-ion concentrations through the terminal voltage model. Figure~\ref{fig:pineapple-validate-data} illustrates three synthetic V-t curves corresponding to different stages of the battery degradation process: early ($\eta_{D_p}=2.5$, $\eta_{D_n}=0.25$, $\eta_{G_p}=2.5$, $\eta_{c_{max,p}}=1$), middle ($\eta_{D_p}=1.5$, $\eta_{D_n}=0.1$, $\eta_{G_p}=3.5$, $\eta_{c_{max,p}}=1$), and late ($\eta_{D_p}=0.3$, $\eta_{D_n}=0.03$, $\eta_{G_p}=3.5$, $\eta_{c_{max,p}}=1$). These variations in SPM and terminal voltage calculation parameters allow us to assess the framework's effectiveness across different degradation states. 

To assess the quality of these solutions, we conduct multiple optimization runs ($n=100$) with different random initializations for each synthetic V-t curve. The results, presented in Figure~\ref{fig:pineapple-validation}, show that while most runs successfully identify a solution that matches the synthetic V-t curves, some fail to converge, leading to higher MSE values.

In addition, we observe a generally strong correlation between MSE and inference error (normalized across four scaling factors), suggesting that MSE can serve as a proxy for inference error to filter out poorly converged solutions (e.g. trapped in suboptimal local minima). Interestingly, this correlation weakens significantly when a 50th percentile MSE threshold is applied to exclude the least plausible solutions. In other words, among solutions with lower MSE values, no clear trend exists between MSE and inference error, indicating that the lowest MSE does not necessarily correspond to the most accurate inferred parameter values. Our results indicate that discrepancies between inferred scaling factors and ground truth (i.e., high inference error) can occur despite achieving a low MSE between the synthetic and LE-PINN-predicted V-t curves. This suggests that the problem is ill-posed, i.e., multiple plausible solutions may exist. This discrepancy may also be a result of small prediction errors introduced by the LE-PINN model.

Using the 50th percentile MSE as a threshold, we summarize the distribution of the remaining plausible solutions and their corresponding V-t curves in Figure~\ref{fig:pineapple-validation}. Notably, the distribution of $\eta_{D_p}$ exhibits greater variability compared to $\eta_{D_n}$. The spread of $\eta_{D_p}$, $\eta_{G_p}$, and $\eta_{c_{max,p}}$ increases from early to late cycles. Despite these variations, the inverse inference distributions encapsulate the ground truth values, and the resulting V-t curves closely align with the synthetic V-t curves, demonstrating the effectiveness of the proposed approach.

%%%%%%%%%%%%%%%%%%%%%%%%%%%%%%%%%%%%%%%%%%%%%%%%%%%% FIG %%%%%%%%%%%%%%%%%%%%%%%%%%%%%%%%%%%%%%%%%%%%%%%%%%%%
\begin{figure}[!hbt]
\centering
\includegraphics[width=1.0\linewidth]{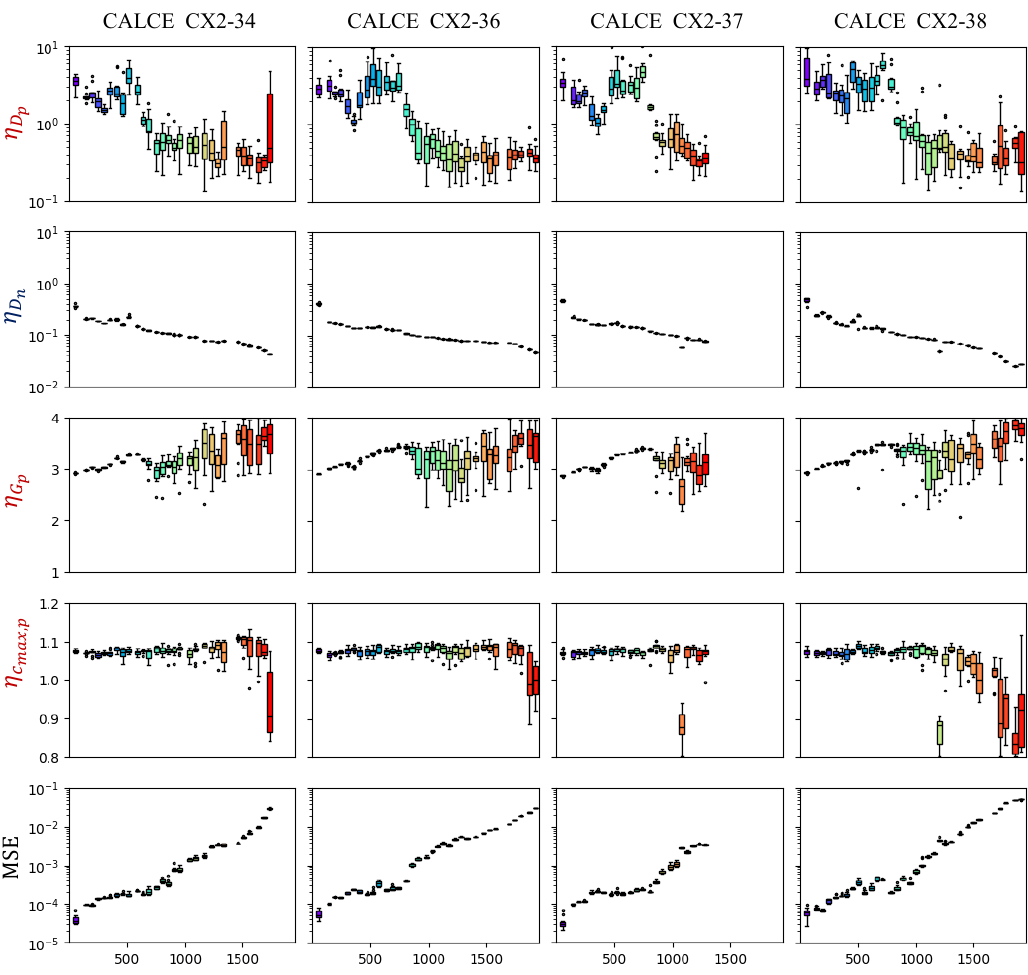} 
\caption{Distribution of inferred parameter values and MSE between measured and predicted V-t curves over cycles during the degradation process for four CX2 batteries. Each distribution is aggregated from 20 independent runs of evolutionary search with different random initializations, for the 50$\%$ of cases with the lowest MSE.}
\label{fig:pineapple-inverse}
\end{figure}

\subsubsection{Real-world datasets.} When applying PINEAPPLE to the estimation of prognostic internal state parameters from the CALCE data, we conduct 20 optimization runs with different random initializations and apply the 50th percentile MSE threshold for each measured V-t curve. This approach provides insights from the ensemble of outcomes. The results for each of the four CX2 battery data are summarized in Figure~\ref{fig:pineapple-inverse}. 

A decreasing trend is generally observed across all four cycle-dependent scaling factors. This aligns with the prior belief that diffusion coefficients in both the cathode and anode progressively decline over the battery’s lifespan, reflecting degradation effects. The maximum concentration value of the battery also decreases at the end of the battery's lifetime, which is characteristic of capacity fade and battery failure.

It is worth noting that the inverse inference performance declines for later cycles, as indicated by the increasing MSE between the measured and predicted V-t curves from LE-PINN. This trend likely reflects the increasing complexity of degradation mechanisms, such as Solid Electrolyte Interphase (SEI) growth, lithium plating, and cathode particle fracture, which cannot be fully captured by a simple SPM framework ~\cite{gopalakrishnan2021composite, han2015simplification, li2018single}. The growing divergence underscores the need for future physics-informed frameworks to embed more advanced multiphysics models of degradation in order to better represent the richness of competing mechanisms during aging. %This decline is potentially attributable to the inherent simplifications in SPM, such as neglecting electrolyte dynamics and spatial heterogeneity, which may limit its ability to fully capture real-world degradation complexities. The mean squared error (MSE) between predicted and measured voltage-time curves was observed to grow approximately logarithmically with cycle number. 
%\clearpage

\subsubsection{Sensitivity analysis of V-t response to SPM parameter perturbations.} To physically interpret the observed variability in inverse inference performance and to better understand the potential ill-posedness of the problem, we perform a local sensitivity analysis of the V-t response with respect to small perturbations in the key SPM scaling factors $\eta_{D_p}$, $\eta_{G_p}$, and $\eta_{D_n}$. This analysis examines how variations in individual parameters influence the V-t trajectory under different operating regimes and provides insight into the robustness and identifiability of the inferred parameters.

\begin{figure}[!hbt]
\centering
\includegraphics[width=1.\linewidth]{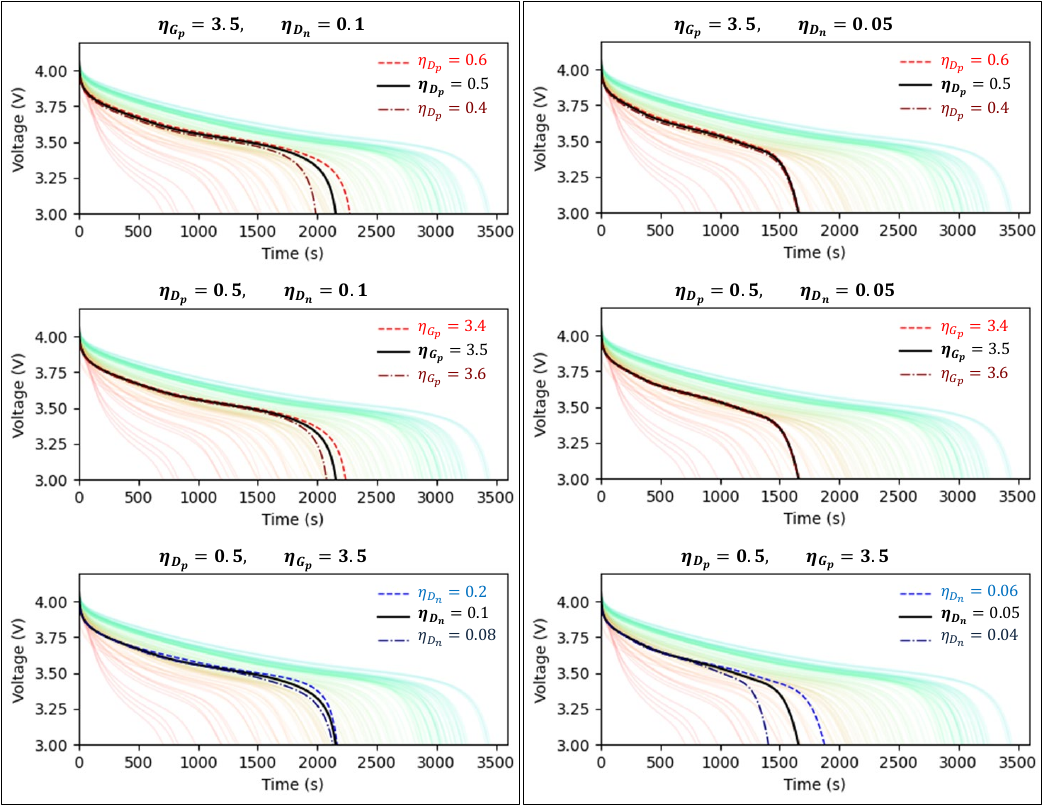} 
\caption{Sensitivity of terminal voltage (V–t) curves to local perturbations in SPM scaling factors. The figure shows the response of the terminal voltage to one-at-a-time perturbations in the scaling factors $\eta_{D_p}$, $\eta_{G_p}$, and $\eta_{D_n}$ under two representative reference states. Left column: reference case with $\eta_{D_p}$=0.5, $\eta_{G_p}$=3.5, and $\eta_{D_n}$=0.1. Right column: reference case with identical positive electrode parameters and a reduced negative electrode diffusion coefficient $\eta_{D_n}$=0.05. In each subplot, the black solid curve denotes the reference V-t trajectory, while dotted curves correspond to small positive and negative perturbations of a single scaling factor, with all other parameters held fixed. To provide physical context, the range of measured V-t curves from the real dataset, as described in Section~\ref{sec:dataset}, is shown in the background as semi-transparent solid lines.}
\label{fig:pineapple-sensitivity}
\end{figure}

Figure~\ref{fig:pineapple-sensitivity} presents the sensitivity results for two representative reference states. In the first case (left column), the reference V–t curve is generated using $\eta_{D_p}$=0.5, $\eta_{G_p}$=3.5, and $\eta_{D_n}$=0.1. In the second case (right column), the positive electrode parameters are kept identical, while the negative electrode diffusion coefficient is reduced to $\eta_{D_n}$=0.05. For each reference state, three subplots are shown, corresponding to one-at-a-time perturbations of $\eta_{D_p}$, $\eta_{G_p}$, and $\eta_{D_n}$, respectively. In the first reference case, the V-t curve exhibits clear sensitivity to small perturbations in the positive electrode parameters $\eta_{D_p}$ and $\eta_{G_p}$, while changes in $\eta_{D_n}$ lead to comparatively minor variations in the V-t curves. This indicates that, when $\eta_{D_n}$ lies in a relatively higher regime, the terminal voltage is significantly influenced by positive electrode dynamics. In contrast, in the second reference case, reducing $\eta_{D_n}$ fundamentally alters the sensitivity profile: the V-t curves become highly sensitive to small perturbations in $\eta_{D_n}$, while exhibiting limited sensitivity to changes in $\eta_{D_p}$ and $\eta_{G_p}$. As a result, multiple combinations of positive electrode parameters can produce nearly indistinguishable voltage trajectories in this regime.

These sensitivity patterns provide explanation for the inverse inference results observed in both the synthetic and real-world datasets (Figures~\ref{fig:pineapple-validation} and~\ref{fig:pineapple-inverse}, respectively). When $\eta_{D_n}$ lies in a higher range, the PINEAPPLE framework exhibits more robust inference of positive electrode scaling factors, as evidenced by smaller variability across inverse inference outcomes. As the battery degrades and $\eta_{D_n}$ decreases into a lower regime characteristic of later cycles, the identifiability of positive electrode parameters diminishes. Consequently, multiple combinations of positive electrode parameters can yield nearly indistinguishable V-t curves, leading to increased variability in the inferred values. These results indicate that the inverse problem exhibits state-dependent identifiability, rather than uniform ill-posedness across the parameter space.

%%%%%%%%%%%%%%%%%%%%%%%%%%%%%%%%%%%%%%%%%%%%%%%%%%%% FIG %%%%%%%%%%%%%%%%%%%%%%%%%%%%%%%%%%%%%%%%%%%%%%%%%%%%

\subsection{Electrochemical scaling factors as indicators of State of Health (SoH) of batteries}

The evolution of inferred electrochemical parameters ($\eta_{D_p}$, $\eta_{D_n}$, $\eta_{G_p}$, $\eta_{c_{max,p}}$) depicted in Figure~\ref{fig:pineapple-inverse} provides insights into the internal state changes during battery cycling and the associated degradation process. 

\subsubsection{Anode Parameters}
Among the four parameters investigated, the negative electrode diffusion coefficient $\eta_{D_n}$ displayed the most consistent decline with increasing cycle number. This steady downward trend indicates a progressive and uniform reduction in lithium-ion transport within the anode as cycling advances. 

This empirical observation is strongly supported by current understanding of battery degradation. Specifically, a dominant degradation mechanism for Li-ion batteries with graphite anodes has been reported to be the continuous growth of the SEI layer, which progressively hinders the transport of Li-ions into the graphite anode, and contributes to capacity fade ~\cite{pinson2012theory}. This is consistent with the observed trend, and this, along with the absence of multiphase transitions and relative uniformity across cells, highlights the potential of $\eta_{D_n}$ as a stable and reliable indicator of SoH and capacity fade. %, where solid state lithium diffusion is progressively hindered by SEI growth and transport limitiations. The observed, consistent decay in inferred $\eta_{D_n}$ across the different cells suggests that graphite anode diffusion is governed primarily by uniform SEI growth and lithium inventory loss, process that dominate gradually and homogeneously throughout the cell's lifetime [Allam \& Onori, 2020]. 

\subsubsection{Cathode Parameters}

In contrast, the positive electrode parameters exhibit much more complex, non-linear behavior as battery cycling progresses. However, there are several interesting observations that can be made from the simultaneous inference of both the diffusion coefficient and geometry parameters.

The diffusion coefficient $\eta_{D_p}$ showed a three-phase evolution with moderate non-linear correlation with increasing cycle number. The three-phase trend observed in $\eta_{D_p}$ may be indicative of sequential degradation mechanisms. The early flat regime is consistent with stable transport and minor electrode degradation. Subsequently, there is a significant mid-life roll-off, which may correspond to transport limitations arising from micro-scale structural degradation, or increased interfacial resistance. The final flat regime may reflect saturation of cathode degradation, with other mechanisms dominating the battery cell's degradation behavior (e.g., particle cracking or structural collapse). Thus, the evolution of $\eta_{D_p}$ potentially serves as a diagnostic fingerprint and window into the dominant degradation mechanisms present in the battery at each stage. 

Similarly, the geometry parameter $\eta_{G_p}$ displayed weak or inconsistent correlation with the number of cycles. However, one interesting observation is that the onset of roll-off in $\eta_{D_p}$ coincided with a sharp increase in variability of $\eta_{G_p}$, suggesting a mechanistic link between cathode transport degradation and structural instability. Since $\eta_{G_p}$ is related to electrode geometry and porosity, its variability likely reflects heterogeneity in micro structural degradation processes, such as pore closure or particle rearrangement ~\cite{Pender2020ElectrodeDegradation, Foster2016BinderDeterioration}. Although $\eta_{G_p}$ is not directly predictive of overall aging, its behavior may still provide qualitative information about the structural evolution of the electrode. 

Lastly, the maximum lithium concentration in the positive electrode $\eta_{c_{max,p}}$ remained largely flat over most of the battery's life and only dropped catastrophically at end-of-life, indicating that it is not a reliable early or mid-life health indicator. The relatively flat profile of $\eta_{c_{max,p}}$ until late-life failure is consistent with the assumption of the SPM model, in which the maximum lithium concentration at the cathode remains effectively constant until severe structural degradation occurs. Such catastrophic decline can be attributed to active material loss, electrolyte depletion, or cathode particle fracture, which typically manifest only in the final phases of battery life ~\cite{Arora1998,Menye2025}. This implies that while $\eta_{c_{max,p}}$ is not sensitive to progressive aging, it may act as a terminal indicator of imminent failure when sudden collapse occurs. 

\begin{figure}[!hbt]
    \centering
    \includegraphics[width=\textwidth]{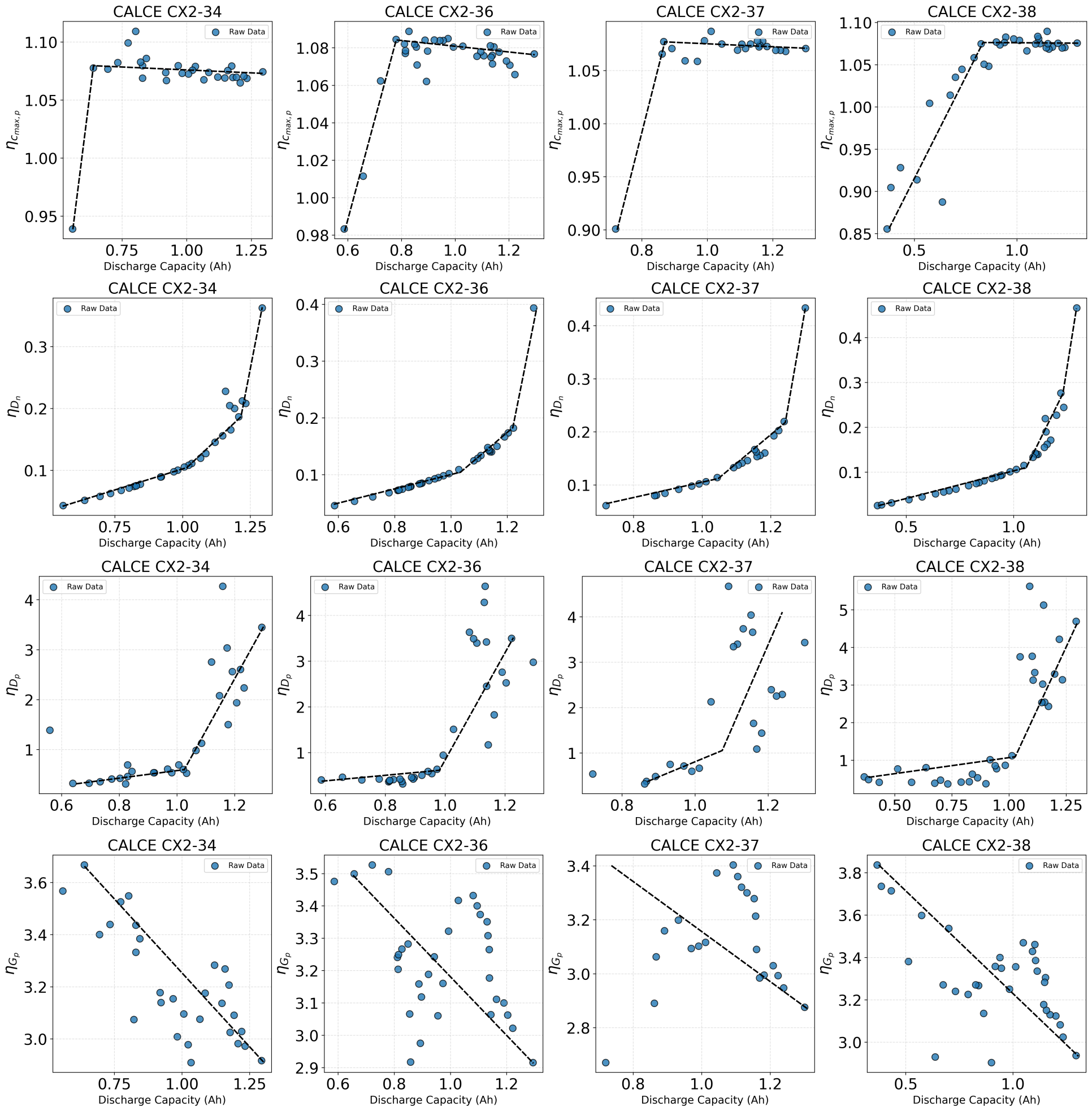}
   \caption{Scatter plots showing the relationship between inferred electrochemical parameters ($\eta_{c_{max,p}}$, $\eta_{D_n}$, $\eta_{D_p}$, $\eta_{G_p}$) and discharge capacity for four batteries (CX2-34, CX2-36, CX2-37, CX2-38).  
    The overlaid trend lines are \textit{manually drawn} to guide qualitative interpretation of parameter evolution and do not represent fitted regression models.}
    \label{fig:regression_all}
\end{figure}

%\break{}

\textbf{Limitations in Analysis.} Overall, this analysis clearly indicates how all four parameters do not evolve in the same way with aging. While $\eta_{D_n}$ evolves monotonically, $\eta_{D_p}$ exhibits a three-phase trajectory, $\eta_{c_{max,p}}$ remains flat until catastrophic failure, and $\eta_{G_p}$ displays high variability. This divergence highlights how different electrochemical processes can dominate at different stages of battery operation, and that combining multiple physical indicators can provide a more comprehensive indicator of battery health. The increasing variance observed in $\eta_{D_p}$, $\eta_{G_p}$, and $\eta_{c_{max,p}}$, but not in $\eta_{D_n}$, suggests that cathode-side degradation processes are increasingly heterogeneous, while anode-side degradation remains relatively uniform and monotonic. 

Nonetheless, care has to be taken in the interpretation of the inferred parameter values. Fundamentally, these parameters are inferred in the context of the SPM model, and while the correlations and trends may seed additional hypothesis and experimental investigation (e.g., the potential link between structural degradation and reduction in transport unveiled by analyzing $\eta_{D_p}$ and $\eta_{G_p}$), certain mechanisms neglected in the model (e.g., the simplifying assumptions about the electrolyte in the SPM) may lead to complexity in the interpretation of the inferred values. These inferred parameter values should ultimately be regarded as effective parameter values which potentially encapsulate variation in multiple actual physical effects (e.g. in the analysis of SEI growth), and it may be important to explicitly resolve key individual physical mechanisms in the final model. Explicit comparison of these (effective) inferred parameters with results from experiments such as EIS or DVA will be particularly insightful in understanding the degree to which these inferred parameters reflect actual experimentally observable degradation. %SEI growth and the effective reduction in anode-side diffusion coefficient $\eta_{D_n}$ which are not explicitly modeled), simplifying assumptions inherent to SPM model can potential errors between the inferred parameters and intrinsic material properties. 

\subsubsection{Correlation between discharge capacity and electrochemical parameters as health indicators}

To complement the temporal evolution analysis, we next performed a direct correlation analysis between discharge capacity and each inferred electrochemical parameter ($\eta_{D_n}$, $\eta_{c_{max,p}}$, $\eta_{D_p}$, $\eta_{G_p}$)  as shown in Figure~\ref{fig:regression_all}.  The discharge capacity is a measure of the battery's capacity, and the comparison of its reduction over time with the evolution of each parameter allows us to mechanistically relate the evolution of each parameter and the battery's aging in a physics-grounded, interpretable way. This can also provide critical insights into their individual suitability as physical indicators of SoH. 

Among the four parameters, the negative-diffusion coefficient $\eta_{D_n}$ and the maximum lithium concentration parameter $\eta_{c_{max,p}}$, display the clearest and most monotonic association with discharge capacity. Their trajectories are smooth and consistent across all cells, indicating that transport limitations in the anode and changes in lithium inventory are strongly coupled to overall capacity loss. These parameters therefore emerge as robust and physically meaningful indicators of degradation that can serve as informative inputs to downstream casual and quantitative health assessment or prognostic models.
%These features therefore emerge as robust, physically meaningful indicators of SoH that can be incorporated directly into predictive maintenance and prognostic models. 

By contrast, the positive electrode diffusion coefficient $\eta_{D_p}$, and the geometric factor $\eta_{G_p}$ exhibit more cell-dependent, non-monotonic patterns. While their trajectories (e.g., onset of roll off) may coincide with structural changes, their weaker and more variable correlations with discharge capacity suggest they are secondary or supporting indicators. They remain useful for capturing cathode-side micro structural effects and may enrich SoH assessment when combined with the stronger anode-related metrics. 

This correlation analysis highlights that different electrochemical process dominate at different stages of battery life. It reinforces the conclusion that a multi-parameter health indicator, as anchored by $\eta_{D_n}$ and $\eta_{c_{max,p}}$ and complemented by $\eta_{G_p}$ and $\eta_{D_p}$, offers a comprehensive and interpretable picture of battery aging, while avoiding assumptions inherent in regression models. 

\subsection{Limitations of SPM model in the PINEAPPLE framework}

The present implementation of PINEAPPLE adopts the Single Particle Model (SPM) as the underlying physical core due to its favorable balance between physical interpretability and computational efficiency. However, this also entails certain model simplifications which are discussed below, such that prediction error may increase due to unresolved physical processes, consistent with the trends observed in Figure~\ref{fig:pineapple-inverse} for later cycles. 

Firstly, the current SPM neglects multiple key physical processes. For example, effects from electrolyte-phase concentration and potential gradients, which can become increasingly important during high C-rate operation and late-stage aging where electrolyte polarization and impedance growth contribute significantly to the measured terminal voltage. As a result, part of the unmodeled electrolyte-induced overpotential may be convolved into effective solid-phase parameters during inverse inference. In addition, while the inferred decline in the anode-side diffusion coefficient is consistent with degradation trends commonly associated with SEI growth, the present framework does not explicitly incorporate interfacial film resistance or loss of cyclable lithium. As such, in an SPM-based formulation without an explicit SEI resistance term, the inferred anode diffusion coefficient should be interpreted as an aggregate health feature that captures combined interfacial and transport losses, rather than as a direct measure of intrinsic bulk diffusion within the graphite particles. Similarly, the present implementation assumes static equilibrium potential functions over the battery lifetime although degradation mechanisms such as loss of active material can shift the accessible stoichiometry window and alter the utilized OCV curve. These effects may also be implicitly absorbed into inferred geometric and capacity-related parameters. 

In addition, the current implementation assumes an isothermal system. While the CALCE dataset used in this study was obtained under a controlled environment near a nominal operating temperature, addition of an explicit thermal model may be required in future extensions as diffusion coefficients and reaction kinetics are strongly temperature-dependent in practice and can significantly influence inverse inference accuracy under varying thermal conditions as may be present in real-world operations. %there exists an inherent identifiability limitation when using the SPM to infer electrochemical parameters solely from voltage–time data.

Importantly, these limitations do not reflect a constraint of the PINEAPPLE framework. Rather, the framework is model-agnostic and scalable by design, allowing alternative battery degradation models such as SPMe or full P2D/DFN formulations—to be embedded as the governing physics when higher fidelity is required. From a computational perspective, the overall complexity of the framework scales with both the number of inferred physical parameters and the optimization depth. For an SPM with approximately five design variables, the combined cost of PINN-based forward evaluation and evolutionary optimization remains lightweight. Scaling to DFN-level models (with its consequent 20 or more parameters) would increase optimization dimensionality and convergence time. Nevertheless, the modular structure of PINEAPPLE enables adaptive trade-offs between physical model fidelity and complexity and computational cost during operation.

Similarly, while the current study focuses on application of PINEAPPLE to LCO-graphite batteries with relatively pronounced voltage signatures, the proposed framework does not depend on specific battery chemistry. The framework is extensible to new battery chemistries such as LFP batteries, although further system validation of the robustness of the inverse inference process to chemistries with different voltage-time signatures is required in future extensions.

\section{Conclusion}

\textbf{Key Findings.} Accurate, real-time estimation of a battery's internal electrochemical state is a critical challenge, with immense potential for the enhancement of next generation BMS. In this work, we presented PINEAPPLE, a framework that combines meta-learned PINNs with an evolutionary search algorithm to estimate multiple physically meaningful hidden internal battery parameters in a computationally tractable manner. This novel framework can help translate voltage measurements into the cycle-dependent evolution of these parameters, solving the ill-posed inverse problem in battery prognostics. Crucially, direct tracking of the evolution of these key parameters is a potential enabler for ``physics-aware'' BMS systems that not just serve as inputs for subsequent SoH/RUL prediction models, but provide a mechanistic, interpretable understanding of why and how the battery is aging. In our proof-of-concept experiments on LCO-Graphite batteries from the CALCE dataset, we indeed observed temporal trends consistent with current understanding of dominant degradation dynamics, i.e. the steady, monotonic decline in negative electrode diffusion coefficient prescribed by continual SEI growth on the anode. %Unlike other approaches in the literature that rely on complex P2D models, PINEAPPLE is built on the simpler Single Particle Model (SPM). This enables direct tracking of the temporal evolution of key parameters such as diffusion coefficients, molar flux, and maximum lithium concentration from voltage–time discharge curves.

\textbf{Future Extensions.} Nonetheless, it is worth acknowledging that the current work remains a proof-of-concept with several avenues for future extensions. 
\begin{enumerate}
    \item First, the SPM is a simplified electrochemical model. Similar to Hassanaly et al.~\cite{hassanaly2024pinn1}, extension of the work to more comprehensive models, such as the P2D framework, can further extend the range of parameters and the richness of the failure mechanisms that can be inferred. 
    \item Second, the current work was focused on a single battery chemistry (LCO-Graphite batteries), with relatively more regular and well-defined charge/discharge cycles. An interesting next step will be to further test the framework under more realistic, stochastic and dynamic load profiles for different battery chemistries (e.g., through the use of other open-source datasets such as NASA, HNEI and Toyota). An important advantage of the framework is that the underlying LE-PINN is mesh-free, meaning it is agnostic to the choice of time points used for prediction and parameter inference. This makes PINEAPPLE extremely versatile, and potentially easily extensible to parameter inference under variable charge and discharge measurement conditions as may be common in practical, real-world usage. 
    \item In particular, digital cell development and design of experiments (DOE) are promising application domains for PINEAPPLE. 
    \item PINEAPPLE can also be scaled to pack-level analysis through a hierarchical Bayesian framework to account for spatio-temporal battery cell inconsistencies at the pack level ~\cite{liionpack2024}.
\end{enumerate} %While the CALCE dataset used in this study was obtained under regular and well-defined charge/discharge cycles, real-world usage is often more irregular, with variations in depth and duration of cycling. The mesh-free nature of LE-PINN prediction allows PINEAPPLE to flexibly extend to parameter inference under such variable charge and discharge conditions, making it suitable for practical deployment.  prior to actual integration and demonstration on a real BMS platform

%Nonetheless, the proposed approach can be extended in several directions. First, it should be tested under dynamic and realistic load conditions and across multiple open-source datasets such as NASA, HNEI, and Toyota to confirm robustness. %Third, the method can be scaled to module and pack levels for monitoring inter-cell variability and predicting remaining useful life (RUL). Finally, demonstrating PINEAPPLE on a real BMS platform would highlight its value for on-the-fly diagnostics and predictive maintenance in both industrial and automotive applications.

Beyond batteries, the core principles of PINEAPPLE, i.e., the versatile combination of physics-informed neural networks with evolutionary optimization for robust parameter inference, can also be applied to similar challenges in other complex physical systems, such as fuel cells, super capacitors, and structural health monitoring of engineered systems, where direct measurement of critical internal states is not feasible. Overall, this framework is potentially applicable to a new generation of models and management systems for critical, engineering assets. 

\section{Acknowledgment}

This research is partly supported by the AME Programmatic: Explainable Physics-based AI for Engineering Modelling \& Design (ePAI) [Award No. A20H5b0142] and RIE2025 Manufacturing, Trade and Connectivity (MTC) Industry Alignment Fund Pre-Positioning (IAF-PP) programme: Battery Remanufacturing for Improved Circular Ecosystems (BRICE) [Award No.: M24N2a0076]. N. Raghavan would like to thank the technical and resource support provided by the IAF-PP Project: Singapore Battery Pack Programme (SGBP2) [Grant No. M23L6a0020].

%\section{Future Work}

%% The Appendices part is started with the command \appendix;
%% appendix sections are then done as normal sections
\appendix

\section{Battery Parameters}
\label{sec:sample:appendix}
Parameters used in the PINEAPPLE framework and associated scaling factors are described in Table~\ref{tab:pineapple-params}.

%%%%%%%%%%%%%%%%%%%%%%%%%%%%%%%%%%%%%%%%%%%%%%%%%%%% TAB %%%%%%%%%%%%%%%%%%%%%%%%%%%%%%%%%%%%%%%%%%%%%%%%%%%%
\begin{table}[htbp]
\centering
\caption{SPM and terminal voltage parameter values.} \label{tab:pineapple-params}
\footnotesize
\begin{tabular*}{\textwidth}{@{\extracolsep{\fill}}>{\raggedright\arraybackslash}p{6cm}>{\raggedright\arraybackslash}p{1cm}>{\raggedleft\arraybackslash}p{1.5cm}>{\raggedleft\arraybackslash}p{1.5cm}>{\raggedright\arraybackslash}p{.5cm}}
    \toprule
    Parameter & Symbol & Scaling factor $^{1}$, $\eta$ & Reference value \\
    \cmidrule{1-5}  
    {\textbf{Lithium Cobalt Oxide (LCO, Cathode)}} \\
    {Diffusion coefficient [m2/s]} & $D_p$ & \cellcolor{lime!20} $1\mathrm{e}^{-1} - 1\mathrm{e}^{1}$ & $3.9\mathrm{e}^{-14}$ & \cite{jiang2022user} \\
    {Geometrical composite coefficient} & $G_p$ & \cellcolor{lime!20} $1 - 4$ & $\frac{R_p}{3\epsilon_pA_pL_p}$ \\
    {Maximum Li concentration [mol/m3]} & $c_{max,p}$ & \cellcolor{lime!20} $0.8 - 1.2$ & $51000$ & \cite{marquis2020long} \\
    {Initial Li concentration [mol/m3]} & $C_p$ & $0.82$ & $30730$ & \cite{marquis2020long} \\
    {Particle radius [m]} & $R_p$ & 5 & $3.0\mathrm{e}^{-6}$ & \cite{marquis2020long} \\
    {Exchange current density [A/m2]} & $j_p$ & 2.5 & $1.0\mathrm{e}^{-1}$ & \cite{swiderska2017kinetics} \\
    {Electrode volume fraction} & $\epsilon_p$ & {} & 0.689 & \cite{almar2019microstructural} \\
    {Electrode area [m2]} & $A_p$ & {} & 0.1 & \cite{munteanu2024single} \\
    {Electrode thickness [m]} & $L_p$ & {} & $7.2\mathrm{e}^{-5}$ & \cite{almar2019microstructural} \\
    \cmidrule{1-5}
    {\textbf{Graphite (Anode)}} \\
    {Diffusion coefficient [m2/s]} & $D_n$ & \cellcolor{lime!20} $1\mathrm{e}^{-2} - 1\mathrm{e}^{1}$ & $3.9\mathrm{e}^{-14}$ & \cite{munteanu2024single} \\
    {Geometrical composite coefficient} & $G_n$ & $2.8$ & $\frac{R_n}{3\epsilon_nA_nL_n}$ \\
    {Maximum Li concentration [mol/m3]} & $c_{max,n}$ & $1$ & $30555$ & \cite{munteanu2024single} \\
    {Initial Li concentration [mol/m3]} & $C_n$ & $1$ & $29866$ & \cite{ai2023improving} \\
    {Particle radius [m]} & $R_n$ & 2 & $5.86\mathrm{e}^{-6}$ & \cite{ai2023improving} \\
    {Exchange current density [A/m2]} & $j_n$ & 3.2 & $7.7\mathrm{e}^{-2}$ & \cite{swiderska2017kinetics} \\
    {Electrode volume fraction} & $\epsilon_n$ & {} & 0.75 & \cite{munteanu2024single} \\
    {Electrode area [m2]} & $A_n$ & {} & 0.1 & \cite{munteanu2024single} \\
    {Electrode thickness [m]} & $L_n$ & {} & $8.3\mathrm{e}^{-5}$ & \cite{munteanu2024single}  \\
    \cmidrule{1-5}
    {\textbf{Common}} \\
    {Current [A]} & $I$ & {} & $1.35$ \\
    {Faraday constant [C/mol]} & $F$ & {} & $96485$ \\
    {Cell temperature [K]} & $K$ & {} & $298.15$ \\ 
    {Universal gas constant [J/mol/K]} & $R_g$ & {} & $8.3145$ \\
    {Internal ohmic resistance [$\Omega$]} & $R_f$ & {0.5} & $0.020$ & \cite{munteanu2024single} \\
    \bottomrule
\end{tabular*}
\raggedright
\footnotesize
$^{1}$ The discharge/charge cycle (degradation) dependent scaling factors are highlighted in color, with their range given.
\end{table}

%% If you have bibdatabase file and want bibtex to generate the
%% bibitems, please use
%%
\clearpage
\bibliographystyle{elsarticle-num} 
\bibliography{cas-refs}

\end{document}